\documentclass[11pt]{article}
\usepackage{comment}
\usepackage{latexsym} 
\usepackage{epsfig,epsf}
\usepackage{graphicx}
\usepackage{amsfonts}
\usepackage{amsmath}
\usepackage{amsthm}
\usepackage{amssymb}
\usepackage{cite}
\usepackage{cancel}

\renewcommand{\theequation}{\thesection.\arabic{equation}}
\newcounter{subequation}[equation]
\makeatletter

\newcommand{\p}{^{\prime}}

\expandafter\let\expandafter\reset@font\csname reset@font\endcsname

\def\subeqnarray{\arraycolsep1pt
    \def\@eqnnum\stepcounter##1{\stepcounter{subequation}%
        {\reset@font\rm(\theequation\alph{subequation})}}
\jot5mm     \eqnarray}

\makeatother

\def\be{\begin{equation}}
\def\lb#1{\label{#1}}
\def\ee{\end{equation}}
\def\bea{\begin{eqnarray}}
\def\eea{\end{eqnarray}}
\def\ba{\begin{array}}
\def\ea{\end{array}}
\def\dd{\partial}

\def\la{\lambda}

\def\ket#1{\left|#1\right>}

\def\one#1{#1^{\raise5pt\hbox{$\scriptstyle\!\!\!\!1$}}\,{}}
\def\two#1{#1^{\raise5pt\hbox{$\scriptstyle\!\!\!\!2$}}\,{}}

\def\tilde{\widetilde}
\def\II{\hbox{{1}\kern-.25em\hbox{l}}}
\def\sang#1#2{\left<{#1} \; {#2} \right>}
\def\ang#1{\left<#1\right>}
\makeatletter
\def\binrel@#1{\begingroup
  \setboxz@h{\thinmuskip0mu
    \medmuskip\m@ne mu\thickmuskip\@ne mu
    \setbox\tw@\hbox{$#1\m@th$}\kern-\wd\tw@
    ${}#1{}\m@th$}%
  \edef\@tempa{\endgroup\let\noexpand\binrel@@
    \ifdim\wdz@<\z@ \mathbin
    \else\ifdim\wdz@>\z@ \mathrel
    \else \relax\fi\fi}%
  \@tempa
}
\let\binrel@@\relax
\def\overset#1#2{\binrel@{#2}%
  \binrel@@{\mathop{\kern\z@#2}\limits^{#1}}}
\def\underset#1#2{\binrel@{#2}%
  \binrel@@{\mathop{\kern\z@#2}\limits_{#1}}}
\makeatother
\newfont{\bbd}{msbm10 scaled\magstep1}

\newcommand{\alphadot}{\stackrel{\cdot}\alpha}

\begin{document}

\begin{titlepage}

\vspace*{1cm}

\begin{center}
{\LARGE \bf{Yang-Baxter operators and scattering amplitudes \\ 
in $\mathcal{N} = 4$ super-Yang-Mills theory}}

\vspace{1cm}

{\large \sf D. Chicherin$^{ab}$\footnote{\sc e-mail:chicherin@pdmi.ras.ru},
S. Derkachov$^{a}$\footnote{\sc e-mail:derkach@pdmi.ras.ru}
and
R. Kirschner$^c$\footnote{\sc e-mail:Roland.Kirschner@itp.uni-leipzig.de} \\
}

\vspace{0.5cm}

\begin{itemize}
\item[$^a$]
{\it St. Petersburg Department of Steklov Mathematical Institute
of Russian Academy of Sciences,
Fontanka 27, 191023 St. Petersburg, Russia}
\item[$^b$]
{\it Chebyshev Laboratory, St.-Petersburg State University,\\
14th Line, 29b, Saint-Petersburg, 199178 Russia}

\item[$^c$]
{\it Institut f\"ur Theoretische
Physik, Universit\"at Leipzig, \\
PF 100 920, D-04009 Leipzig, Germany}
\end{itemize}
\end{center}
\vspace{0.5cm}

\begin{abstract}
Yangian symmetry of amplitudes in $\mathcal{N}=4$ super Yang-Mills theory
is formulated in terms of eigenvalue relations for monodromy matrix
operators. The Quantum Inverse Scattering Method provides the appropriate
tools to treat the extended symmetry and to recover as its consequences 
many known features like cyclic and inversion symmetry, BCFW recursion,
Inverse Soft Limit construction, Grassmannian integral representation, 
$\mathrm{R}$-invariants and  on-shell diagram approach.
\end{abstract}

\end{titlepage}


\newpage

{\small \tableofcontents}
\renewcommand{\refname}{References.}
\renewcommand{\thefootnote}{\arabic{footnote}}
\setcounter{footnote}{0} \setcounter{equation}{0}

\section{Introduction}
In the weak coupling context
dual superconformal symmetry of scattering
amplitudes in super Yang-Mills theory at large $N$ has been discovered in~\cite{DHKS08} 
and analyzed  in \cite{Br08}.
Relying on the gauge/string duality the relation of amplitudes to light-like
Wilson loops has been estabished earlier in \cite{AM}. Here the
superconformal symmetry of the Wilson loop is important in the calculation
of amplitudes at strong coupling. The relation of both superconformal
symmetries has  been understood by string T-duality \cite{BRTW,BM}. The
integrability of the related sigma model has been used in strong coupling
amplitude computations \cite{AMSV}.  
Yangian symmetry
has been established as the unification of both kinds of superconformal
symmetries~\cite{Dr09}. The extended symmetries became part of
the modern treatment  and understanding of gauge theories and their
relation to strings. The Yangian symmetry of amplitudes appeared in the
first instant as the elegant and compact formulation of the symmetry
properties of  known amplitude expressions and their relation to Wilson
loop expectation values. Applications of Yangian symmetry of amplitudes have
been studied e.g. in \cite{KS,DF}. It is desirable to use the extended symmetries as
tools of calculation. 
We propose  a formulation which  allows to exploit easily all the advantages
of  Yangian symmetry in calculations and investigations of amplitudes. 

In most investigations of this symmetry on amplitudes Drinfeld's formulation
in terms of algebra generators~\cite{Drin85} has been applied. 
In the case of the symmetry
algebras $g\ell(N)$ a formulation in the framework of the
Quantum Inverse Scattering Method (QISM) was worked out earlier by 
L.D.~Faddeev and collaborators~\cite{Fad,FST,TaFa,KulSk,KRS,Skl91}. 
We rely on the advantages of the QISM formulation here. It has features
appearing natural to physicists, because it emerged as the mathematical
formulation of the methods developed for the Heisenberg spin chain and other
integrable models.
Our considerations are heavily based on the  ideas and techniques of QISM.  
In particular this means that we associate with a considered $n$-particle
amplitude $M$ a spin chain with $n$ sites.

We are going to formulate  the condition of Yangian symmetry of
amplitudes $M$ following
\cite{ChK13} in terms of the monodromy matrix $\mathrm{T}(u)$ as
the eigenvalue relation
\be \label{TM0}
 \mathrm{T}(u) \ M = C\ M. 
\ee 
The eigenvalues $C$ play an auxiliary role here. 

The monodromy matrix is an ordered matrix product of $\mathrm{L}$-operators each 
referring to one site of the spin chain. The $\mathrm{L}$-operator is an operator-valued 
matrix with elements composed out of the 
symmetry algebra generators in the relevant representation. 
In our case the representation space is the one corresponding to the single 
particle states of the vector multiplet of the 
$\mathcal{N}=4$ extended SYM including their helicities and momenta. 

The $\mathrm{L}$-operator depends on the spectral parameter $u$ and in
general the monodromy matrix depends on $u_1, ..., u_n$. The above symmetry 
condition refers to the homogeneous case of coinciding spectral parameters.

In \cite{ChK13} considering $g\ell(N)$-symmetric spin
chains 
 the eigenvalue problem for {\it inhomogeneous quantum monodromy} 
matrices has been formulated as the Yangian symmetry condition.  Eigenfunctions of the 
monodromy called
Yangian symmetric correlators have been constructed and some relations implied by the
symmetry condition have been derived.

The Yang-Baxter $\mathrm{R}$-operator\footnote{
The notion of $\mathrm{R}$-operator is not to be confused with the one of
$\mathrm{R}$-invariants~\cite{DHKS08}.} 
is another important ingredient of QISM.
It acts in the tensor product of two infinite-dimensional local quantum spaces and intertwines
a pair of $2$-site monodromies, i.e. two representations of the Yangian algebra.
The corresponding intertwining relation is known as Yang-Baxter $\mathrm{RLL}$-relation.
We shall use Yang-Baxter $\mathrm{R}$-operators to generate more solutions
of the symmetry condition from given ones. This works under the condition
that the corresponding  $\mathrm{R}$-operators can be permuted with the monodromy
matrix in the eigenvalue relation.
In this perspective   it has been proposed to consider 
{\it generalized Yang-Baxter} relations~\cite{ChK13} 
because higher point eigenfunctions of the monodromy define as  kernels
the corresponding generalized Yang-Baxter operators obeying these relations.  

In this paper we are going to apply the methods developed in~\cite{ChK13} 
to the $g\ell(4|4)$-symmetric spin chain.
This particular integrable quantum-mechanical system has played a crucial role  in 
unravelling the integrable structures of $\mathcal{N} = 4$ super-Yang-Mills theory 
in composite operator renormalization~\cite{Beis} and is the relevant one
here.
The comparison to the more general case considered in \cite{ChK13} illustrates
the specifics of the situation of super-Yang-Mills field theory. 
The presentation of the present paper is kept basically 
self-contained with respect to the  details of the previous one. 

Our main statement is the following: 
The Yangian symmetry can be formulated as the 
 the eigenvalue problem for the monodromy matrix 
and in solving it we recover
the crucial constructions for SYM scattering amplitudes 
such as the link integral representation,
the Inverse Soft Limit (ISL) construction, on-shell diagrams 
which have been 
pioneered by Arkani-Hamed and collaborators~\cite{ABCCK09,AH09,ACC09,AH12,AH10}
and also $\mathrm{R}$-invariants~\cite{DHKS08}.
These concepts have been developed originally without  reference to Yangian symmetry 
or integrable structure. 
We emphasize that all these structures arise inevitably from the basic concepts of QISM.
The only input for us is the appropriate eigenvalue problem for monodromy
matrices which we solve exploiting  solely concepts and constructions typical for QISM.

We shall show that the Yangian symmetry condition is compatible with the
iterative BCFW construction of amplitudes and that the elementary
three-particle amplitudes are Yangian symmetric. 

However, there  are eigenfunctions of the Yangian symmetry condition
more elementary rather than those three-particle amplitudes, called {\it basic
states}. They are formed by  products of delta functions 
of spinor variables each referring to one site of 
the $n$-site spin chain. 
The local structure of these states implies absence of interactions. 
 
It is   remarkable  that the amplitude terms can be obtained by acting
on such basic states by  products of Yang-Baxter $\mathrm{R}$-operators defined from the 
$\mathrm{L}$ matrices by the $\mathrm{RLL}$ intertwining relation.
The $\mathrm{R}$-operators act bilocally  touching just two sites of the spin chain.
The sequential action by Yang-Baxter $\mathrm{R}$-operators on the basic state 
results in more and more entangled, nonlocal solutions. 
The representation of amplitudes in terms of operator actions has been
found earlier in~\cite{MasSk09} without noticing the connection to Yang-Baxter
relations.
We shall also show the relation to the Inverse Soft Limit (ISL) construction.
Representing the  $\mathrm{R}$-operator in the form of a  contour integral 
 a sequence of $\mathrm{R}$-operator actions transforms into the 
 Grassmannian  
link integral representation~\cite{ABCCK09}.  

The eigenvalue problem for the monodromy is invariant with respect to cyclic
shifts of spin chain sites
and it transforms in a simple manner with respect to reflection of the site ordering. 
The fact that a sequence
of $\mathrm{R}$-operators generates an eigenstate is based on the possibility 
to pull the sequence through the monodromy matrix.
The latter is provided by cyclicity, reflection and the $\mathrm{RLL}$-relation.

In order to recover on-shell diagrams from the perspectives of QISM 
we consider integral operators in spinor-helicity variables whose kernels are
eigenfunctions of the monodromy. 
In this way we identify the Yang-Baxter $\mathrm{R}$-operator,  the basic
tool of our construction, as the  integral operator with the cut $4$-point
amplitude as kernel.

We shall also consider the eigenvalue problem for the monodromy 
in super momentum twistor variables introduced by Hodges~\cite{Hod09}. 
The corresponding construction of eigenfunctions follows the previous pattern. 
The basic state has a local form and it is formed as a product of delta functions of 
super momentum twistors 
or identity each referring to one site of the spin chain. 
The other eigenfunctions of the monodromy 
are generated  again by acting on the basic state with a sequence of bilocal $\mathrm{R}$-operators 
now in super momentum twistor variables. We recover $\mathrm{R}$-invariants in this way.

For all constructions relevant for SYM scattering amplitudes 
 we can restrict ourselves to the case of the homogeneous monodromy which is obtained 
from the inhomogeneous one by taking all 
spectral parameters equal. Then the $\mathrm{R}$-operators appear with argument value
zero. 
Solving the eigenvalue condition with inhomogeneous monodromy matrices
we obtain spectral parameter dependent deformations of the amplitude
expressions. The deformation affects the  dilatation weights, i.e. the
superconformal Casimir of the representation, shifting them  off the physical value 
originally determined by the scattering particles. 
Keeping the parameters at
generic values  provides advantages related to analytic continuations.
 Parameter deformed amplitudes  have been considered and  the use for regularization has
been pointed out in~\cite{FLMPS12}. We show that our techniques
allows to obtain easily deformed amplitude expressions and discuss their
applications.

The plan of the paper is as follows.  
In Sect.~\ref{sec-LR} we introduce the basic notations and objects relevant for QISM
such as the $\mathrm{L}$-operator, the monodromy matrix and the $\mathrm{R}$-operator 
in the case of spinor-helicity variables 
for the $g\ell(4|4)$ symmetry algebra recalling the framework from~\cite{ChK13}.
In Sect.~\ref{sec-BCFW} we show that the BCFW recurrent procedure is 
compatible with Yangian symmetry  where we understand the latter from the point 
of view of QISM as the monodromy eigenvalue condition.
We also represent $3$-point amplitudes in terms of $\mathrm{R}$-operators 
acting on a basic state.   
In Sect.~\ref{sec-cycl} we prove that the eigenvalue relation for the 
homogeneous monodromy is invariant with
respect to cyclic shift of the spin chain sites and reflection of the site ordering.
In Sect.~\ref{sec-ISL} we establish the connection with the ISL construction 
of scattering amplitudes.
In Sect.~\ref{sec-canon} we discuss canonical transformations which relate 
the present construction with the 
one  in~\cite{ChK13}. 
In Sect.~\ref{sec-NH} we discuss eigenvalue problems for inhomogeneous monodromy
matrices.
We construct $3$- and $4$-point eigenfunctions by a sequence 
of operators acting on basic states.
In Sect.~\ref{sec-Int} we recall the connection between the 
eigenvalue problem for the inhomogeneous monodromy
and the generalized Yang-Baxter relation and construct 
integral Yang-Baxter operators whose kernels 
are eigenfunctions of the monodromy.     
We show in Sect.~\ref{sec-MT} that the $\mathrm{R}$-invariants appearing 
beyond the MHV level are
recovered by Yang-Baxter operator action on appropriate basic states in
momentum twistor variables.

\section{L-operator, R-operator and Yangian symmetry}
\setcounter{equation}{0}
\label{sec-LR}

Here we introduce the basic tools needed in our construction.
We start with two sets of mutually conjugate variables\footnote{$\mathbf{p}$ is not to be confused with
a momentum of a
scattering particle and $\mathbf{x}$ is not to be confused with a region momentum.}
$\mathbf{x} = (x_{a})_{a=1}^{N+M}$ and $\mathbf{p}=(p_{a})_{a=1}^{N+M}$
where the index $a$ enumerates $N$ bosonic components ($a=1,\cdots,N$) and $M$ fermionic components ($a=
N+1,\cdots,N+M$).
These variables respect canonical commutation relations with the graded commutator $\{x_a,p_b] = - \delta_{ab}$,
i.e. commutation relation for bosons ($a,b = 1, \cdots, N$) and anticommutation relation for fermions ($a,b = N+1,\cdots, N+M$).
Later we shall restrict our discussion to $N=4$ bosons and $M=4$ fermions, i.e. $4|4$,
as it is the relevant case for $\mathcal{N}=4$ SYM.
We are interested in Jordan-Schwinger type representations of the symmetry algebra
$g\ell(N|M)$
whose generators $x_a p_{b}$  can be unified in a matrix and 
 supplemented with a {\it spectral parameter} $u$ term proportional to the unit
matrix,
\be \lb{L}
\mathrm{L}(u) = u + \mathbf{x} \otimes \mathbf{p},
\ee
or more explicitly in component notations,
$$
\mathrm{L}_{ab}(u) = u \,\delta_{ab} + x_a \,p_b\,.
$$
This matrix is referred to as \underline{$\mathrm{L}$-operator}.
It is easy to check that it satisfies the {\it fundamental commutation relation}, called $\mathcal{R}\mathrm{LL}$-relation,
\be \lb{FCR}
\mathcal{R}_{ab,ef}(u-v) \, \mathrm{L}_{ec}(u)\,
\mathrm{L}_{fd}(v) = \mathrm{L}_{bf}(v)\,
\mathrm{L}_{ae}(u)\, \mathcal{R}_{ef,cd}(u-v),
\ee
with Yang's $\mathcal{R}$-matrix, 
$ \mathcal{R}(u) = u + \mathrm{P}$, where $\mathrm{P}$ is the graded permutation 
and $a,b, \cdots = 1,\cdots,N+M$. 
The latter equation for the $\mathrm{L}$-operator is equivalent to the 
defining (anti)commutation relations of $g\ell(N|M)$.

One of the merits of the Quantum Inverse Scattering Method~\cite{Fad,TaFa,FST,KulSk,Skl91,KRS} is that it enables us to construct
involved nonlocal objects out of local ones, where the interaction of  several copies of the
introduced degrees of freedom is included in integrable way.
 Pursuing this strategy
we consider $n$ copies of canonical variables ($\mathbf{x}_1, \cdots,
\mathbf{x}_{n}$) and ($\mathbf{p}_1, \cdots, \mathbf{p}_{n}$)
which are interpreted as the dynamical variables of a quantum spin chain with $n$ sites, 
i.e. $\mathbf{x}_i$ and $\mathbf{p}_i$
are local variables of the $i$-th site.
Further we construct the {\it homogeneous monodromy} matrix $\mathrm{T}(u)$ of
the $n$-site  chain as the ordered matrix product of
$n$ 
$\mathrm{L}$-operators each referring to one site of the spin chain,
\be \lb{T}
\mathrm{T}(u) = \mathrm{L}_{1} (u)\, \mathrm{L}_{2} (u) \cdots \mathrm{L}_{n} (u) = 
\begin{array}{c}\includegraphics{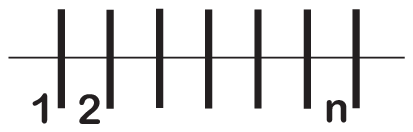} \end{array}
\ee
Here $\mathrm{L}_{i} (u) $ is the $\mathrm{L}$-operator (\ref{L}) with
$x_a, p_a$ substituted by the local canonical pairs at site $i$, $x_{i,a},
p_{i,a}$.   
It is easy to understand \cite{Fad}  
that the highly nonlocal monodromy matrix satisfies
the fundamental commutation relation (\ref{FCR}) too, which is also known as
the Yangian relations.
In Sect.~\ref{sec-NH} we shall consider a more general situation of inhomogeneous
monodromy  which depends on $n$ spectral parameters.

In applications to SYM the Yangian algebra has been  usually used in 
Drinfeld's formulation~\cite{Drin85} working with generators $J^{0}$ and $J^{1}$, where
$J^{0}$ generate $g\ell(N|M)$. The equivalence of the latter  formulation to
the QISM formulation used here is well known and has been
explained in detail in \cite{Molev}.
$\mathrm{T}(u) $ is the generating function of the Yangian algebra
generators in a particular representation,
\be \lb{J}
\mathrm{T}_{ab}(u) =  \sum^{n-1}_{m=-1} u^{n-m-1} J^{m}_{ab}.
\ee
$J^{0}_{ab}$ and $ J^{1}_{ab}$ represent the generators in Drinfeld's
formulation; we have  $J^{-1}= I$ and the other ingredients  $ J^{m}_{ab}, m>1,$
are determined from the generators. 
The fundamental  $\mathcal{R}\mathrm{LL}$-relation (\ref{FCR}) 
implies the Yangian algebra relations, i.e. the commutation relations of
$J^{0}_{ab}$ and $ J^{1}_{ab}$ and the Serre relations. It implies also that
 the higher level  $J^{m}$, $m>1$, 
and the commutation relations between them are consequences of
the relations involving  the lower two levels $m= 0,1$ only.

One may redefine the basis of generators by
\be \label{redef} 
J^{(0)}_{ab} = J^{0}_{ab} - \alpha \delta_{ab} \sum J^{0}_{cc}, \ \ 
J^{(1)}_{ab} = J^{1}_{ab} - \beta \sum J^{0}_{ac}J^{0}_{cb} 
\ee
 with arbitrary   
constants $\alpha$ and $\beta$. This does not change the algebra.

 We formulate the condition of Yangian symmetry, applicable in
particular to SYM scattering amplitudes specifying (\ref{TM0}), 
as the eigenvalue relation with the monodromy operator (\ref{T}),
\be \label{eigenu} 
\mathrm{T}_{ab}(u) \,
M(\mathbf{x}_1,\cdots,\mathbf{x}_N) = C(u) \,\delta_{ab}\, M(\mathbf{x}_1,\cdots,\mathbf{x}_N) \,.
\ee
The eigenvalue $C$ depends on the spectral parameter $u$.

In our case  $\mathrm{T}_{ab}(u)$ is constructed according to (\ref{T})
with the $\mathrm{L}$ operators of the form (\ref{L}). This implies that 
$J^{0}_{ab}$ are sums of the local generators of the symmetry algebra
$g\ell(N|M)$ of the spin chain, and $J^{1}_{ab}$ are bilocal generators
$$
J^0_{ab} = \sum_{1 \leq i \leq n} x_{a,i} \,p_{b,i}\;,\;\;\;
J^1_{ab} = \sum_{1 \leq i < j \leq n} x_{a,i} \,p_{c,i} \,x_{c,j} \,p_{b,j}\,.
$$
The eigenvalue condition in terms of $\mathrm{T}_{ab}(u)$  (\ref{eigenu})
implies by  the decomposition (\ref{J}) the following conditions in terms of the Drinfeld
generators, 
$$J^0_{ab} M = C_0  \delta_{ab} M,  \ \  J^1_{ab} M = C_1   \delta_{ab} M.  $$  
$C_0 $ and $ C_1$ appear in the expansion of $C(u)$ as
$C(u) = u^n ( 1 + C_0 u^{-1} + C_1 u^{-2} + ...)$. 
The eigenvalue conditions involving the higher level $J^{m}$, $m>1$,
are consequences of the  latter ones  
because the higher level operators are obtained from the
ones on the first two levels. By the above redefinition (\ref{redef}) 
with appropriate parameters $\alpha, \beta , 
\alpha = \frac{1}{N+M}, \beta = \frac{C_1}{C_0^2}$ 
the symmetry condition can be cast into the form  
$$ J^{(0)}_{ab} \, M = 0,\ \ \  J^{(1)}_{ab} \, M  = 0 $$
as it appeared in the first papers on Yangian symmetry of amplitudes.

Being a generating function is not the main point for preferring the
monodromy matrix. More important are its composition from local building
blocks of the spin chain and its connection to Yang-Baxter relations of
several types. 

As a further  tool of the QISM 
we introduce the \underline{$\mathrm{R}$-operator} by means of the intertwining {\it $\mathrm{RLL}$-relation},
\be \lb{RLL}
\mathrm{R}_{12}(u-v)\, \mathrm{L}_1(u) \,\mathrm{L}_2(v) = \mathrm{L}_1(v)\, \mathrm{L}_2(u) \,\mathrm{R}_{12}(u-v)\,.
\ee
As indicated by the subscripts  the operator $\mathrm{R}_{12}(u-v)$ acts nontrivially in two sites $1,2$ of the spin chain 
and as the result permutes the  spectral parameters of the involved $\mathrm{L}$-operators. We can also say that $\mathrm{R}$
is an intertwining operator that intertwines a pair of 
representations, $\mathrm{L}_1(u)\mathrm{L}_2(v)$ and
$\mathrm{L}_1(v)\mathrm{L}_2(u)$, of the Yangian algebra defined 
by the fundamental commutation relation (\ref{FCR}).
In the case of our interest to be specified below the $\mathrm{RLL}$-relation (\ref{RLL}) can be depicted as follows
\be \lb{RLLpic}
\begin{array}{c}
\includegraphics{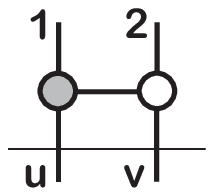}
\end{array} =
\begin{array}{c}
\includegraphics{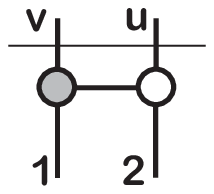}
\end{array}
\ee
It will be explained further in Sect.~\ref{sec-Int}.
 To prevent confusion we add the remark that this Yang-Baxter $\mathrm{RLL}$-relation
(\ref{RLL}) differs from the fundamental Yang-Baxter relation (\ref{FCR}). They are different
representations of a general algebraic relation.
 In (\ref{RLL}) the $\mathrm{L}$ operators enter in matrix product. They act on different
spaces indicated by the subscripts $1,2$. The operator $\mathrm{R}_{12}$
 acts on the tensor product of these two spaces and it is not a matrix in our case. 
 In (\ref{FCR}) both $\mathrm{L}$ act on the same space. They
enter in matrix tensor product (expressed by explicit indices). $\mathcal{R}$
is a $(N+M)^2 \times (N+M)^2$ matrix. 

The equation (\ref{RLL}) can be taken as the defining condition of the
$\mathrm{R}$-operator and formal
algebraic operations lead to the solution~\cite{K13}
\be \lb{R}
\mathrm{R}_{12}(u) = \Gamma(u) \,(\mathbf{p}_1 \cdot \mathbf{x}_2)^{-u}
=\int\limits^{\infty}_{0} \frac{\mathrm{d} z}{z^{1-u}} e^{-z (\mathbf{p}_1 \cdot \mathbf{x}_2)}
= \frac{i}{2 \sin \pi u } \int_{\mathcal{C}}
\frac{\mathrm{d} z}{z^{1-u}} e^{-z (\mathbf{p}_1 \cdot \mathbf{x}_2)}
\ee
where we use the shorthand notation for the inner product $(\mathbf{p}_1 \cdot \mathbf{x}_2) = p_{a,1} x_{a,2}$
and the contour $\mathcal{C}$ encircles clockwise the positive real
semi-axis starting at $+\infty - i \epsilon$, surrounding $0$, and ending at $+\infty + i\epsilon$.

In \cite{ChK13}  explicit expressions of $2$,$3$,$4$,$5$-point
symmetric correlators and  
$n$-point correlators for a particular configuration are given.
 Further  
 an iterative procedure which allows to
construct higher point eigenfunctions has been proposed.  Transformations
which relate eigenfunctions for different $n$-site monodromies have been
discussed. 
In particular it has been  shown that the eigenvalue problems for the
monodromy
with cyclically shifted or reflected labels of spin chain sites 
is essentially equivalent to the initial one.
The one-dimensional structure of the associated spin chain 
is reflected in the {\it cyclicity} property of the eigenfunctions.
In the  construction the  $2$-point correlators have played a special role and
they turned out to be 
a basic tool for constructing higher-point correlators. 
They have been   identified with the particular 
Yang-Baxter $\mathrm{R}$-operator proposed in~\cite{K13}.
This kind of Yang-Baxter operators is related to one of the factors of
the Yang-Baxter operator  in $s\ell(N)$  
and plays a role in Baxter operator constructions \cite{DKK}.

Up to now the formulation is  rather general. We are going to specify
the dynamical variables $\mathbf{x}$, $\mathbf{p}$
for application to scattering amplitudes in $\mathcal{N}=4$ SYM.
We are interested in two types of variables: 
spinor helicity variables (see next Subsection) 
and super momentum twistors (see Sect.~\ref{sec-MT}).

\subsection{Spinor helicity variables}
\label{sec-hel}
The  external particle states of  the colour-stripped $\mathcal{N}=4$ SYM scattering
amplitudes  can be parameterized 
by  a light-like momentum $p$ (i.e. $p^2 = 0$) which factorizes in a pair of spinors 
of opposite helicities
$p = \lambda \otimes \tilde{\lambda}$, i.e. 
$p_{\alpha \dot{\alpha}} = \lambda_{\alpha} \tilde{\lambda}_{\dot{\alpha}}$\,, 
and by particle type and helicity. The latter information is encoded
by a 
polynomial in the Grassmann variables $\eta_{A}$ ($A=1,\cdots,4$).
Therefore we take as local dynamical variables of the related spin chain the
following specifications,
$\mathbf{x} \to \begin{pmatrix} \lambda_\alpha , & \partial_{\tilde{\lambda}_{\dot{\alpha}}} , & \partial_{\eta_A} \end{pmatrix}$ and
$\mathbf{p} \to \begin{pmatrix} \partial_{\lambda_\alpha} , & -\tilde{\lambda}_{\dot{\alpha}} , & - \eta_A \end{pmatrix}$.
Then the $\mathrm{L}$-operator (\ref{L}) acquires the form
\be \lb{Lsh}
\mathrm{L}(u) =
\begin{pmatrix}
u\cdot 1 + \lambda \otimes \partial_{\lambda} & -\lambda \otimes \tilde{\lambda} & -\lambda \otimes \eta \\
\partial_{\tilde{\lambda}} \otimes \partial_{\lambda} & u\cdot 1 - \partial_{\tilde{\lambda}} \otimes \tilde{\lambda}
& - \partial_{\tilde{\lambda}} \otimes \eta \\
\partial_{\eta} \otimes \partial_{\lambda} & - \partial_{\eta} \otimes \tilde{\lambda} & u\cdot 1 - \partial_{\eta} \otimes \eta
\end{pmatrix}.
\ee
One sees that it is a matrix with operator elements being generators 
of the superconformal algebra in super spinor variables~\cite{Wit03}.
The monodromy matrix (\ref{T}) defines the Yangian algebra.
The action of the corresponding $\mathrm{R}$-operator (\ref{R}) on a function $F$ produces
the BCFW shift,
\be \lb{Rhel}
\mathrm{R}_{ij}(u) F(\lambda_i,\tilde{\lambda}_{i},\eta_{i}|\lambda_j,\tilde{\lambda}_{j},\eta_{j}) =
\int\frac{\mathrm{d}z}{z^{1-u}}
F(\lambda_{i}-z \lambda_{j}, \tilde{\lambda}_i,\eta_i | \lambda_j , \tilde{\lambda}_j + z \tilde{\lambda}_i , \eta_j + z \eta_i )\,.
\ee
Now we have identified the spin chain dynamical variables with the variables describing 
external states of the scattering amplitude, and the Yangian symmetry statement for amplitudes is
translated into the eigenvalue relation for the monodromy (\ref{eigenu}). 


\section{BCFW and Yangian symmetry}
\setcounter{equation}{0}
\label{sec-BCFW}

In~\cite{Br08} it has been checked that the BCFW recursion relations are compatible with
the dual superconformal symmetry.
We are going to show how the monodromy condition (\ref{eigenu}) can be applied to
check the  Yangian symmetry of tree scattering amplitudes and leading singularities of 
loop corrections.
The known proofs of this fact are either based on the explicit form of the tree amplitudes~\cite{Dr09}, 
i.e. on explicit solutions
of the BCFW recursion \cite{DrHe08}, or exploit the Grassmannian formulation~\cite{AH10}. 
We would like to understand the Yangian symmetry directly from
the BCFW recursion relations without solving them. We shall show that the procedure of BCFW iteration is
compatible with the Yangian symmetry: the building blocks, the
$3$-point amplitudes, obey the Yangian symmetry condition and the BCFW
construction preserves the symmetry.  
Let us emphasize once more  that we
rely here only on the  
monodromy matrix formulation~(\ref{eigenu}) of the Yangian symmetry.

\begin{figure}
\begin{center}
\includegraphics{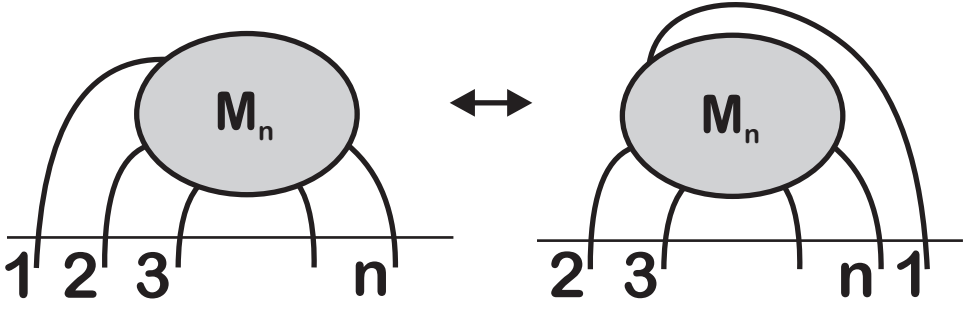}
\caption{\label{figCycl} The monodromy matrix (\ref{T}) is symmetric with respect to cyclic shift of 
spin chain sites (legs of the on-shell diagram) on the space of its eigenfunctions.}
\end{center}
\end{figure}

To set up the notations we start with recalling the BCFW relations~\cite{BCF04,BCFW05} in their 
supersymmetrised version~\cite{AH08,Br08,EFK09}.
Consider the colour-stripped scattering amplitude of $n$ particles in ${\mathcal N}=4$ SYM  
$$
M_n = \sum_{k} M_{k,n} = \delta^4\left(\sum_{i=1}^{n} p_i \right) \mathcal{M}_n \;,\;\;\;
\mathcal{M}_n = \sum_{k} \mathcal{M}_{k,n}.
$$
Here the four-dimensional delta function takes  momentum conservation into
account. 
$\mathcal{M}_{k,n}$ (as well as $M_{k,n}$) has degree $4k$ in Grassmann 
variables $\eta^{A}_i$ ($A=1,\cdots,4$)
specifying the external states. Each momentum $p_i$ is light-like $p_i^2 =0$ and factorizes into two spinors
$p_i = \lambda_i \otimes \tilde{\lambda}_i$.
The relation to the notation referring to the helicity violation 
 is $M_{k,n} = \mathrm{N}^{k-2}\mathrm{MHV}_n$ and $M_{1,3} = \overline{\mathrm{MHV}}_3$.
It is known that $\mathcal{M}_{k,n}$ and $M_{k,n}$ are cyclically symmetric. 
How cyclic symmetry follows in our approach relying on  the monodromy matrix~(\ref{T})
will be discussed in Sect.~\ref{sec-cycl} (see Fig.~\ref{figCycl}).

The supersymmetric BCFW relation \cite{AH08} has the form
\be \lb{BCFW}
\mathcal{M}_{k,n} = \sum_{L,R} \int \mathrm{d}^4 \eta \,
\mathcal{M}_L \bigl( \eta_1\,,\,\lambda_1(z_*)\,,\,\tilde{\lambda}_1\,;\, \eta\,,\, - P_{1\cdots i}(z_*) \bigr)
\frac{1}{P^2_{1\cdots i}}\,
\mathcal{M}_R \bigl( \eta_n(z_*)\,,\,\lambda_n\,,\, \tilde{\lambda}_n(z_*)\,;\, \eta\,,\, P_{1\cdots i}(z_*) \bigr)
\ee
where the amplitudes $\mathcal{M}_L$ and $\mathcal{M}_R$ have $i+1$ and $n-i+1$ legs respectively,
the total Grassmann degree of $\mathcal{M}_L$ and $\mathcal{M}_R$ is four units
larger than the one of $\mathcal{M}$. Here the dependence of $\mathcal{M}_L$ and 
$\mathcal{M}_R$ on their spinor and Grassmann arguments $\hat i =
(\lambda_i, \tilde \lambda_i, \eta_i)$ is displayed only partially, emphasizing 
those active in the considered relation. Further, the variables
$\widehat {i+1}$ in $\mathcal{M}_L$ and  $\hat n$ in $\mathcal{M}_R$ are 
expressed in terms of the intermediate momentum $P$.     
The following standard notations for the BCFW shift are used
\be \lb{BCFWshift}
\lambda_1(z) = \lambda_1 - z \lambda_n \;,\;\;\;
\tilde{\lambda}_n(z) = \tilde{\lambda}_n + z \tilde{\lambda}_1 \;,\;\;\;
\eta_n(z) = \eta_n + z \eta_1\,,
\ee
$$
z_* = \frac{P^2_{1\cdots i}}{\langle n|P_{1\cdots i} |1 ]} \;,\;\;\;
P_{1\cdots i}(z) = P_{1\cdots i} - z \lambda_n \tilde{\lambda}_1 \;,\;\;\;
P_{1\cdots i} = p_1 + p_2 + \cdots + p_i\,.
$$
The statement that we are to prove is that the amplitude $M_{k,n}$
(with the momentum conserving delta function included)
is an eigenfunction of the monodromy matrix (\ref{T})
\be \lb{eigen}
\mathrm{T}(u) \, M_{k,n} = u^k(u-1)^{n-k} \cdot M_{k,n} \,.
\ee
Consequently the amplitude is an eigenfunction of the generators of Yangian algebra~(\ref{J}),
$$
J^m_{ab} \, M_{k,n} = \delta_{ab}\, \frac{(-)^{m+1}(n-k)!}{(m+1)!(n-k-m-1)!}
\cdot M_{k,n}\,.
$$

We have explained above (\ref{redef}) that  
one may redefine the basis of generators by
 $J^{(0)}_{ab} = J^{0}_{ab} - \alpha \delta_{ab} \sum J^{0}_{cc}, \ \ 
J^{(1)}_{ab} = J^{1}_{ab} - \beta \sum J^{0}_{ac}J^{0}_{cb} $ with arbitrary   
constants $\alpha$ and $\beta$. In this example we have  the explicit form
of the eigenvalues $C(u)$ and thus of the expansion coefficients $C_0, C_1$.
Then with the particular choice of 
$ \alpha= \frac{1}{N+M}, \beta= \frac{n-k-1}{2 (n-k)}$
the symmetry condition can be cast into the form of the invariance conditions
$ J^{(0)}_{ab} \, M = 0,\ \ \  J^{(1)}_{ab} \, M  = 0 $.

\subsection{Symmetry of the convolution}

First we consider one term in the sum (\ref{BCFW}) and prove that after
multiplying by the momentum delta function it obeys
(\ref{eigen}) if the involved $M_L, M_R$ obey the corresponding monodromy
eigenvalue relations.  Then our argument proceeds by induction in the number
of legs.

 We deform the  BCFW relations (\ref{BCFW}) by
 substituting propagators $\frac{1}{P^2} \to \frac{1}{(P^2)^{1-\Delta}}$ in a special manner.
Indeed let us define $M_n(\Delta)$ by the recurrence relation
\be \lb{MD'}
M_{k,n}(\Delta) =
\int \mathrm{d}^4 \eta \,
\int\limits^{\infty}_{0} \frac{\mathrm{d} z}{z^{1-\Delta}} \int \mathrm{d}^4 P_0 \, \delta(P_0^2)\,
M_L \bigl( \eta_1\,,\,\lambda_1(z)\,,\, \tilde{\lambda}_1\,;\, \eta\,,\, - P_0 \bigr)
M_R \bigl( \eta_n(z)\,,\,\lambda_n\,, \,\tilde{\lambda}_n(z)\,;\, \eta\,,\, P_0 \bigr).
\ee
We emphasize that $M_L$ and $M_R$ do contain the momentum conservation delta function.
The analogous formula at $\Delta = 0$ has been used in \cite{AH09,MasSk09} to rewrite BCFW in twistor space.
We integrate easily 
 over $P_0$ and $z$ because these variables enter via  
the energy-momentum delta function contained in the product of the amplitudes,
\be \lb{propagator}
\int\limits^{\infty}_{0} \frac{\mathrm{d} z}{z^{1-\Delta}} \int \mathrm{d}^4 P_0 \, \delta(P_0^2)\,
\delta^4 (P_{1\cdots i}-P_0 - z \,\lambda_n \tilde{\lambda}_1) =
\frac{\langle n|P_{1\cdots i} |1 ]^{-\Delta}}{\left(P^2_{1\cdots i}\right)^{1-\Delta}}
\equiv \frac{1}{\Pi^2_i(\Delta)}\,.
\ee
There are two ways to make the previous expression well-defined. 
The first one appeals to split signature $(2,2)$ of space-time such that all 
spinors are real.
Then we can assume that $\langle n|P_{1\cdots i} |1 ] > 0$, $P^2_{1\cdots i} >0$
for the integration of the delta function over $z$ in (\ref{propagator}) 
to be well defined.
The second possibility is to consider complexified momenta and to interpret the delta function 
in (\ref{propagator}) according to Dolbeault. For more details see for example~\cite{MaSk09}.
Thus (\ref{MD'}) takes the form
$$
\mathcal{M}_{k,n}(\Delta) =
\int \mathrm{d}^4 \eta \,
\mathcal{M}_L \bigl( \eta_1\,,\,\lambda_1(z_*)\,,\, \tilde{\lambda}_1\,;\, \eta\,,\, - P_{1\cdots i}(z_*) \bigr)
\frac{1}{\Pi^2_i(\Delta)}\,
\mathcal{M}_R \bigl( \eta_n(z_*)\,,\,\lambda_n\,,\, \tilde{\lambda}_n(z_*)\,;\, \eta\,,\, P_{1\cdots i}(z_*) \bigr)
$$
that is a slight modification of one term in the BCFW sum (\ref{BCFW}).
At $\Delta \to 0$ the standard BCFW relation arises, $M_{k,n}(\Delta=0) = M_{k,n}$.
Further we represent the BCFW shifts by differential operators
$$
\begin{array}{lll}
M_L \bigl( \eta_1\,,\,\lambda_1(z)\,,\, \tilde{\lambda}_1\,;\, \eta\,,\, - P_0 \bigr) &=&
e^{-z \lambda_n \partial_{\lambda_1}} M_L \bigl( \eta_1\,,\,\lambda_1\,, \,\tilde{\lambda}_1\,; \,\eta\,,\, - P_0 \bigr) \,,\\ [0.2 cm]
M_R \bigl( \eta_n(z)\,,\,\lambda_n\,, \,\tilde{\lambda}_n(z)\,; \,\eta\,,\, P_0 \bigr) &=&
e^{z \tilde{\lambda}_1 \partial_{\tilde{\lambda}_n} + z \eta_1 \partial_{\eta_n}}
M_R \bigl( \eta_n\,,\,\lambda_n\,,\, \tilde{\lambda}_n\,;\, \eta\,,\, P_0
\bigr).
\end{array}
$$
 In view of (\ref{R}) (or (\ref{Rhel}))
this allows to  rewrite BCFW by means of the 
$\mathrm{R}$-operator which acts on the $1$-st and $n$-th legs of the amplitude,
\be \lb{MD}
M_{k,n}(\Delta) =
\mathrm{R}_{1 n}(\Delta) 
\int \mathrm{d}^4 \eta_0 \,
\mathrm{d}^4 P_0 \, \delta(P_0^2)\,
M_L \bigl( \eta_1\,,\,\lambda_1\,,\, \tilde{\lambda}_1\,;\, \eta_0\,,\, - P_0 \bigr)\,
M_R \bigl( \eta_n\,,\,\lambda_n\,,\, \tilde{\lambda}_n\,; \,\eta_0\,, \,P_0 \bigr).
\ee
We are going to calculate the action of the monodromy matrix on the amplitude
\be \lb{TM}
\mathrm{L}_{n-1} (u) \cdots \mathrm{L}_{2} (u) \mathrm{L}_{1} (u-\Delta) \mathrm{L}_{n} (u)
\, M(\Delta) \;\;\;\; \text{at} \;\; \Delta \to 0\,.
\ee
Here we consider an inhomogeneous monodromy matrix, i.e. we  introduce
the regularization parameter $\Delta$ 
in the monodromy matrix (\ref{T}). We shall show in Sect.~\ref{sec-cycl}
that the orderings of the spin chain sites 
in (\ref{T}) and (\ref{TM}) are equivalent from the point of view of the monodromy
eigenvalue condition. 
This can be also understood taking into account the cyclic symmetry of 
the color-stripped amplitude as a fact (known or to be proven separately as
in Sect. 4). 
It is clear that the previous relation does not contain singularities at $\Delta \to 0$ and
we can freely substitute $\Delta = 0$ in the monodromy matrix and in the amplitude $M(\Delta)$.
We notice that in spite of the apparent pole in (\ref{R}) at $u = 0$ the operator 
$\mathrm{R}_{12}(u=0)$
is finite on the space of distributions we deal with. We will see this 
below on explicit examples. 

Then we specify the assumption of the induction that the amplitude with a number of
external particles lower than $n$ is an eigenfunction of the monodromy (\ref{T}) with an eigenvalue $C_m$
\be \lb{asump}
\mathrm{L}_m(u) \cdots \mathrm{L}_1(u) \, M_m = C_m \cdot M_m\;,\;\;\; \text{at} \;\; m < n\,.
\ee
Now we act by the  monodromy matrix (\ref{TM}) on $M(\Delta)$ (\ref{MD}) suppressing integrations for a while
$$
\mathrm{L}_{n-1} (u) \cdots \mathrm{L}_{2} (u) \mathrm{L}_{1} (u-\Delta) \mathrm{L}_{n} (u) \, \mathrm{R}_{1 n}(\Delta) \,M_L \,M_R =
$$
\be \lb{TRM}
=\mathrm{R}_{1 n}(\Delta) \, \mathrm{L}_{n-1} (u) \cdots \mathrm{L}_{i+1} (u) \,
\underline{\mathrm{L}_{i} (u) \cdots \mathrm{L}_{1} (u) \, M_L} \,\mathrm{L}_{n} (u-\Delta) \,M_R\,.
\ee
In the previous formula we have been allowed  to pull the $\mathrm{R}$-operator through the monodromy due to 
the $\mathrm{RLL}$-relation (\ref{RLL}) that is a key observation.
To simplify the underlined factor we use the assumption of induction in the
form (\ref{asump})
\be \lb{TM=L-1M}
\mathrm{L}_{i} (u) \cdots \mathrm{L}_{1} (u) \, M_L = C_L \cdot\mathrm{L}^{-1}_{0} (u) \, M_L\,,
\ee
i.e. $M_L$ having $i+1$ legs is an eigenfunction of the monodromy.

In order to invert the $\mathrm{L}$-operator (\ref{L}) we note that
\be \lb{LLcas}
\mathrm{L}(u)\, \mathrm{L}(v) = u v + (u+v-1+ \mathrm{c}  ) \, \mathbf{x} \otimes \mathbf{p}
\ee
where  in the considered case of spinor helicity variables (\ref{Lsh})
\be \lb{cas}
\mathrm{c} \equiv \left( \mathbf{p} \cdot \mathbf{x} \right) = 2 + \lambda\dd_{\lambda} - \tilde{\lambda}\dd_{\tilde{\lambda}}-\eta\dd_{\eta}
\ee
is the {\it Casimir} operator of the superconformal algebra characterizing the chosen representation.
It commutes with the $\mathrm{L}$-operator, $[\mathrm{L}(u)\,,\,\mathrm{c}] =0$.
The Casimir operator is related to the  helicity operator $h$ as $h = 1-\frac{\mathrm{c}}{2}$. For
$\mathcal{N}=4$ SYM amplitudes $h = 1$ and we have $\mathrm{c}=0$. This simplifies considerably all the following
calculations with $\mathrm{L}$-operators. 
Consequently at $v= 1-u$  (\ref{LLcas}) takes the form
\be \lb{inv}
u(1-u) \,\mathrm{L}^{-1}(u) = \mathrm{L}(1-u)\,.
\ee
Further we define the transposed $\mathrm{L}$-operator integrating by parts
$$
\int \mathrm{d}^4 \eta \int \mathrm{d}^4 P \, \delta(P^2) \left[\mathrm{L}(u) \Phi \right] \Psi =
\int \mathrm{d}^4 \eta \int \mathrm{d}^4 P \, \delta(P^2) \,\Phi \left[ \mathrm{L}^T(u) \Psi \right]
$$
where the functions $\Phi(P,\eta)$ and $\Psi(P,\eta)$ are even in the  Grassmann 
variables $\eta_A$.
It is easy to check that
\be \lb{transp}
\mathrm{L}^T(u) =  -\mathrm{L}(1-u)\,,
\ee
and taking  (\ref{inv}) and (\ref{transp}) together we have
\be \lb{-1T}
u(u-1) \, \mathrm{L}^{-1 T}(u) = \mathrm{L}(u)\,.
\ee
Thus we see that matrix inversion and operator transposition reproduce the $\mathrm{L}$-operator
with the  initial dependence on the spectral parameter. This is indispensable for the Yangian symmetry to hold
in the form (\ref{eigenu}) and
 is due to the vanishing of the Casimir operator $\mathrm{c}$ on the space of amplitudes. 
In the case of nonzero Casimir operator
the $M_{k,n}$ can be an eigenfunction of the inhomogeneous monodromy matrix
(depending on a set of $n$ arbitrary spectral parameters)
only if the propagator has the nonstandard form (\ref{propagator}) with nonzero $\Delta$
that does not admit a direct field theory interpretation. This case will be addressed in Sect.~\ref{sec-NH}.

Thus substituting (\ref{TM=L-1M}) in (\ref{TRM}), taking into account integrations in (\ref{MD}) and
integrating by parts by means of (\ref{-1T}) one obtains
$$
\frac{C_L}{u(u-1)}\, \mathrm{R}_{1 n}(\Delta) \int \mathrm{d}^4 \eta_0
\, \mathrm{d}^4 P_0 \, \delta(P_0^2)\,
M_L \, \underline{\mathrm{L}_{n-1} (u) \cdots \mathrm{L}_{i+1} (u)\, \mathrm{L}_{0} (u)\, \mathrm{L}_{n}(u-\Delta) \, M_R}\,.
$$
Then due to the induction assumption (\ref{asump}) we conclude that the underlined factor is equal
to $C_R \,M_R + \mathcal{O}(\Delta)$, and taking $\Delta \to 0$
we obtain that the monodromy matrix applied to a particular term $M$ of the full 
amplitude,
$\mathrm{L}_{n-1} (u) \cdots \mathrm{L}_{2} (u)\, 
\mathrm{L}_{1} (u)\, \mathrm{L}_{n} (u) \, M$, 
results in
\be \lb{eigenv}
\frac{C_L C_R}{u(u-1)}
\int\limits^{\infty}_{0} \frac{\mathrm{d} z}{z} \int \mathrm{d}^4 \eta_0 \, \mathrm{d}^4 P_0 \, \delta(P_0^2)\,
M_L \bigl( \eta_1\,,\,\lambda_1(z)\,,\, \tilde{\lambda}_1\,;\, \eta\,,\, - P_0 \bigr)\,
M_R \bigl( \eta_n(z)\,,\,\lambda_n\,,\, \tilde{\lambda}_n(z)\,;\, \eta\,, \,P_0 \bigr).
\ee
Thus each term of the BCFW sum is an eigenfunction of the monodromy with eigenvalue
$\frac{C_L C_R}{u(u-1)}$.
Let us remind that each term of the BCFW sum is a residue of the contour integral over
a Grassmannian.
Thus we have shown that the residues are Yangian invariant, i.e. they are eigenfunctions of the monodromy.
It remains  to check that this eigenvalue is the same for all
terms and thus  $M_{k,n}$ is also an eigenfunction of the monodromy.
Stating the other way: all residues of the contour integral correspond to the same eigenvalue.
To prove it we  first check that the $3$-point amplitudes are eigenfunctions of the monodromy
as the starting point  of the induction, and then by means of the BCFW iteration  we
calculate the  eigenvalues for all tree amplitudes and leading singularities.
\subsection{Three-point amplitudes}
\label{sec-3point}
The basis of BCFW recursion are the $3$-point MHV and anti-MHV amplitudes,
\begin{eqnarray} 
\lb{MHV3}
\mathrm{M}_{2,3}(p_1,p_2,p_3) &=& \frac{\delta^{4}(p_1+p_2+p_3)\delta^{8}(\lambda_1 \eta_1 + \lambda_2 \eta_2 + \lambda_3 \eta_3)}
{\ang{12} \ang{23} \ang{31}} \,,
\\[0.2 cm]
\lb{aMHV3}
\mathrm{M}_{1,3}(p_1,p_2,p_3) &=& 
\frac{\delta^{4}(p_1+p_2+p_3)\delta^{4}([1 2] \eta_3 + [2 3] \eta_1 + [3 1] \eta_2)}{[ 1 2 ] [ 2 3 ] [3 1] } \,,
\end{eqnarray}
out of which BCFW reconstructs arbitrary tree amplitudes.
Exploiting this idea we will check first that they respect Yangian symmetry, i.e. they are eigenfunction of the $3$-site monodromy, 
and then we extend it to all tree amplitudes and leading singularities by means of BCFW.

In order to represent the $3$-point amplitudes in a convenient form 
we follow the general strategy of Quantum Inverse Scattering Method
constructing complicated nonlocal objects out of local ones.
We start with the direct product of trivial local states which we refer to as
the {\it basic state} $\Omega_{k,n}$,
\be \lb{vac}
\Omega_{1,3} = \delta^2(\lambda_1) \, \delta^2(\lambda_2) \,\delta^{2}(\tilde{\lambda}_3)\delta^{4}(\eta_3)\,
,
\ee
and act on it by $\mathrm{R}$-operator (\ref{Rhel}) two times obtaining nonlocal expression 
for the $3$-point anti-MHV amplitude (\ref{MHV3}),
\be \lb{opaMHV3} 
\mathrm{M}_{1,3}(p_1,p_2,p_3) =
\mathrm{R}_{12}\,\mathrm{R}_{23} \,\Omega_{1,3} \,.
\ee
Here and in the following we use the short-hand notation for the $\mathrm{R}$-operator
$\mathrm{R}_{ij} \equiv \mathrm{R}_{ij}(0)$ taken at zero spectral parameter. 
We see that the  bilocal 
$\mathrm{R}$-operator (\ref{R}) generates just the nontrivial interactions relevant 
for the super Yang-Mills theory.

We have the  analogous situation  for the $3$-point MHV amplitude
\be \lb{opMHV3}
\mathrm{M}_{2,3}(p_1,p_2,p_3) = 
\mathrm{R}_{23}\, \mathrm{R}_{12} \,\Omega_{2,3} \,\,,\,\,\, 
\Omega_{2,3} = \delta^2(\lambda_1) \,\delta^{2}(\tilde{\lambda}_2) \delta^{4}(\eta_2)\, \delta^{2}(\tilde{\lambda}_3) \delta^{4}(\eta_3)\,.
\ee
At the end of this Subsection we check the latter formula. 

Using formulae (\ref{opaMHV3}), (\ref{opMHV3}) it is straightforward to check that the
supersymmetric three-point amplitudes (\ref{MHV3}), (\ref{aMHV3}) are eigenfunctions
of the monodromy matrix of three-site spin chain and to calculate corresponding eigenvalues.
Indeed, we take into account the explicit  form of the $\mathrm{L}$-operator (\ref{Lsh}) and
obtain immediately how it acts on delta functions of spinors which are local basic states
\be \lb{Lvac}
\mathrm{L}(u) \, \delta^{2}(\lambda) = (u-1) \cdot \delta^{2}(\lambda) \;,\;\;\;
\mathrm{L}(u) \, \delta^{2}(\tilde{\lambda})\delta^{4}(\eta) = u \cdot \delta^{2}(\tilde{\lambda})\delta^{4}(\eta)\,.
\ee
Consequently the basic state $\Omega_{2,3}$ (\ref{opMHV3}) formed as the 
direct product of local basic states
is an eigenstate of the $3$-site monodromy,
$$
\mathrm{L}_1(u)\mathrm{L}_2(u)\mathrm{L}_3(u) \,\Omega_{2,3} = u^2 (u-1)\cdot\Omega_{2,3}\,.
$$
Then in view of (\ref{opMHV3}) and the operator intertwining $\mathrm{RLL}$-relation (\ref{RLL})
we obtain that the $3$-point MHV amplitude (\ref{MHV3}) is an eigenfunction as well with the same eigenvalue, 
$$
\mathrm{L}_1(u)\mathrm{L}_2(u)\mathrm{L}_3(u) \, M_{2,3} =
\mathrm{R}_{23} \mathrm{R}_{12} \, \mathrm{L}_1(u)\mathrm{L}_2(u)\mathrm{L}_3(u)
\, \Omega_{2,3} =
$$
$$
= (u-1)u^2 \cdot\mathrm{R}_{23} \mathrm{R}_{12} \,\Omega_{2,3} =
(u-1)u^2 \cdot M_{2,3}\,.
$$
Complementing the latter calculation with the one for anti-$\mathrm{MHV}_3$ amplitude (\ref{aMHV3}) we 
obtain a pair of relations which constitute the basis of the induction proof started in the previous Subsection,
\be \lb{vert}
\begin{array}{c}
\mathrm{L}_1(u)\mathrm{L}_2(u)\mathrm{L}_3(u) \,M_{2,3} = (u-1) u^2 \cdot M_{2,3}\,, \\ [0.2 cm]
\mathrm{L}_1(u)\mathrm{L}_2(u)\mathrm{L}_3(u) \,M_{1,3} = u (u-1)^2 \cdot M_{1,3}\,.
\end{array}
\ee
Let us note that  (\ref{opaMHV3}), (\ref{opMHV3}) are actually well known. 
In~\cite{MasSk09} the scattering amplitudes in super-twistor variables 
have been represented in a form similar to (\ref{opaMHV3}) as a sequence of operators of Hilbert transformations
acting on delta functions of super-twistors. If we stay in the spinor-helicity notations then 
the formulae (\ref{opaMHV3}), (\ref{opMHV3})
correspond exactly to on-shell diagrams of Arkani-Hamed et al~\cite{AH12}. 
Further comments on this point will be given in Sect.~\ref{sec-Int}.

The nontrivial part of our statement is that these amplitude constructions can be extracted solely 
in the framework of Quantum Inverse Scattering Method solving an eigenvalue problem for the monodromy (\ref{eigenu})
without resorting to any auxiliary concepts or assumptions.

In order to demonstrate (\ref{opaMHV3}), (\ref{opMHV3}) we prove  first
the representation for the anti-$\mathrm{MHV}_3$ amplitude.
It will be convenient for us here and in the following to adopt the shorthand notation 
$\delta^{2|4}(\tilde{\lambda}) \equiv \delta^{2}(\tilde{\lambda}) \delta^{4}(\eta)$
which is rather natural since $\tilde{\lambda}$ and $\eta$ are subjected to identical BCFW shifts (\ref{BCFWshift}).
Taking into account (\ref{Rhel}) we have
$$
\mathrm{R}_{23} \, \delta^{2}(\lambda_2) \,\delta^{2|4}(\tilde{\lambda}_3) =
\int \frac{\mathrm{d}z}{z} \,\delta^2(\lambda_3-z\lambda_3)\, \delta^{2|4}(\tilde{\lambda}_3+z \tilde{\lambda}_2) =
$$
$$
=\frac{[12]}{[13]} \,\delta([23])\,
\delta^{2}\left(\lambda_2+ \frac{[31]}{[21]} \lambda_1 \right)\,
\delta^{4}\left(\eta_3+ \frac{[31]}{[12]} \eta_2 \right),
$$
where we have rewritten one of the delta functions as
$\delta^2(\tilde\lambda_3+z \tilde\lambda_2) = [1 2] \delta([23]) \delta( [31]+z [21])$.
We admit that this representation for delta function is not completely satisfactory since 
it does not allow to fix the sign unambiguously. This representation is 
in the spirit of the paper~\cite{AH11} 
where a delta function is substituted by an analytic function with a simple pole and 
corresponding integrals are calculated by means of Cauchy's theorem.
Here and further in similar calculations
we apply the formal rule and do not pay attention 
to the overall sign which is not very important since we are interested in eigenfunctions.
Then we denote $p = p_1+p_2+p_3$, $q = q_1 + q_2 + q_3$ and apply once more
an $\mathrm{R}$-operator
$$
\mathrm{R}_{12} \mathrm{R}_{23} \,\Omega_{1,3} =
\frac{[12]}{[13]} \int\frac{\mathrm{d}z}{z} \delta([23]+z [13])
\delta^{2}(\lambda_1-z\lambda_2)\,
\delta^{2}\left( \frac{p\,|1]}{[21]} \right)
\delta^{4}\left(\eta_3 + \frac{[31]}{[12]} (\eta_2+z\eta_1) \right) =
$$
$$
= \frac{[12]}{[23] [31]}
\delta^{2}\left( \frac{p\,|3]}{[13]} \right)
\delta^{2}\left( \frac{p\,|1]}{[21]} \right)
\delta^{4}\left(\frac{q}{[12]} \right) =
\frac{\delta^4(p)\delta^4([12]\eta_3 + \text{cycl})}{[12] [2 3] [31]}\,.
$$

This calculation clearly demonstrates that the $\mathrm{R}$-operator (\ref{Rhel}) 
at vanishing spectral parameter argument is well defined 
on the space of distributions we deal with 
because their support does not contain the point $z = 0$.
In Sect.~\ref{sec-NH} we consider inhomogeneous monodromies and construct their eigenfunctions 
which allows to keep the argument of the $\mathrm{R}$-operator at nonzero values. 
The present formulae  follow from those 
in the limit of  vanishing $\mathrm{R}$-operator arguments that is equivalent
to taking  all spectral parameters of the monodromy equal.

The formulae (\ref{opaMHV3}), (\ref{opMHV3}) imply that the amplitudes can be constructed acting by $\mathrm{R}$-operators on the
basic state formed by delta functions of spinors.
They demonstrate that  amplitudes which have rather nonlocal forms can be represented in fact
as a sequence of operators each touching only two sites of the periodic spin chain applied to
the basic state $\Omega_{k,n}$.
It resembles very much the Algebraic Bethe Ansatz diagonalization 
of the quantum spin chain and 
the Separation of Variables method~\cite{Skl91}.

\subsection{All eigenvalues}
Having obtained the eigenvalues for the three point amplitudes we are ready to calculate
the eigenvalues for arbitrary tree amplitudes.
The BCFW recursion for $\mathrm{N}^{k-2}\mathrm{MHV}$ amplitude $M_{k,n}$ can be represented symbolically as \cite{AH08}
\be \lb{BCFWsymb}
M_{k,n} = M_{1,3} \otimes M_{k,n-1} + \sum_{i = 2}^{k-3} \sum_{m=3}^{n-1} M_{i,m} \otimes M_{k-i+1,n-m+2}\,.
\ee
This formula specifies the Grassmann degrees of the terms in (\ref{BCFW}). It says that
the amplitudes of degree $4k$ with $n$ legs are
constructed out of amplitudes with lower numbers of legs and lower degree in
Grassmann variables 
resulting in the 
 inductive construction of tree amplitudes with respect to $k$ and $n$.
Applying the eigenvalue relation (\ref{eigenv}) according to the pattern (\ref{BCFWsymb}) it is easy to check (\ref{eigen}),
$$
\mathrm{L}_1(u)\mathrm{L}_2(u)\cdots\mathrm{L}_n(u)\cdot M_{k,n} = u^{k}(u-1)^{n-k} \,M_{k,n} \,.
$$
Indeed according to (\ref{eigenv}) $M_{1,3} \otimes M_{k,n-1}$ is an eigenfunction of the monodromy with eigenvalue
$$
\frac{u(u-1)^2 \cdot u^{k}(u-1)^{n-k-1}}{u(u-1)} = u^{k}(u-1)^{n-k}\,,
$$
and $M_{i,m} \otimes M_{k-i+1,n-m+2}$ corresponds to the eigenvalue
$$
\frac{u^{i}(u-1)^{m-i} \cdot u^{k-i+1}(u-1)^{n-m+i-k+1}}{u(u-1)} = u^{k}(u-1)^{n-k}\,.
$$
Thus each term in BCFW sum is an eigenfunction of the monodromy corresponding to the same
eigenvalue and consequently the amplitude $M_{k,n}$ as well.
Finally, the Yangian symmetry relation (\ref{eigen}) is proven and the eigenvalues of
the monodromy matrix are calculated.

For on-shell diagrams including loops the BCFW iteration has been formulated
in~\cite{AH10,AH12}. The above arguments can be adapted easily to include
the terms involving the contributions from cut loop propagators. The
induction is to be set up to go first up in the number of legs at fixed
maximal loop order and then proceed to the next loop level. 
 Leading singularities are eigenfunctions of the monodromy matrix as well.
Furthermore, the corresponding eigenvalues are fixed by $k$ and $n$, thus they
are the same as for tree amplitudes (see (\ref{eigen})).

Notice that we have proven actually that any linear combination of the
 terms in the BCFW sum is Yangian symmetric. The symmetry condition does not
fix the particular one appearing as the physical amplitude.
In the Grassmannian approach to scattering amplitudes~\cite{ABCCK09} the BCFW terms 
of the tree 
amplitudes $\mathrm{M}_{k,n}$ and leading singularities of its loop corrections 
are identified with residues
of the contour integral over Grassmannian $G(k,n)$. Thus such a contour integral is 
an eigenfunction
of the monodromy with eigenvalue $u^k(u-1)^{n-k}$ (see (\ref{eigen})).

\section{Reflection and cyclicity}
\setcounter{equation}{0}
\label{sec-cycl}

The eigenvalue relation (\ref{eigenu}), i.e. the Yangian symmetry statement, 
allows for reflection and  cyclic shift transformations of spin chain sites 
which looks especially simple in the considered case of $g\ell(4|4)$ spin chain relevant for 
$\mathcal{N} = 4$ SYM.
Actually for nonzero values of Casimir operator (\ref{cas}) or
the other symmetry algebras the cyclic permutation leads to inhomogeneous
shifts of the spectral parameters in some $\mathrm{L}$-operators
and a change in the eigenvalue \cite{ChK13}. We will address the case of 
nonzero Casimir operators
in Sect.~\ref{sec-NH}.

The {\it reflection} property appears by multiplication of the eigenvalue relation
by the inverse of the monodromy matrix. The latter is calculated from the
inversion relation for the $\mathrm{L}$-operators (\ref{inv}).
\be \lb{reflect}
\mathrm{L}_1(u) \cdots \mathrm{L}_n(u) \, M(1,\cdots,n) = C \, M(1,\cdots,n) \Rightarrow
\mathrm{L}_{n}(1-u) \cdots \mathrm{L}_{1}(1-u) \, M(1,\cdots,n) = C\p \, M(1,\cdots,n)
\ee
where $C\p = C^{-1} u^n (1-u)^n $. 

Apparently a pair of monodromy matrices (\ref{T}) with cyclically shifted sites
are not related to each other in a simple way. 
Actually these operators are different.
However their eigenvalue problems  are equivalent. 
Now we demonstrate without reference to our previous results
that in the  special case of $g\ell(4|4)$-symmetric spin chain
the monodromy matrix (\ref{T}) acts on its eigenfunctions in {\it cyclically} symmetric
way, i.e.
\be \lb{cyc}
\mathrm{L}_1(u) \cdots \mathrm{L}_n(u) \, M(1,\cdots,n) = C \, M(1,\cdots,n) \Rightarrow
\mathrm{L}_{\sigma_1}(u) \cdots \mathrm{L}_{\sigma_n}(u) \, M(1,\cdots,n) = C \, M(1,\cdots,n)
\ee
where $\sigma_1,\cdots,\sigma_n$ is a cyclic permutation of $1,2,\cdots,n$.

Let us define the graded matrix transposition of the matrix $K$ 
with operator-valued entries
(bosonic as well as fermionic) by
$(K^t)_{ab} = (-)^{\bar{a}\bar{b}} K_{ba}$, where $\bar{a}$ denotes the  
Grassmann degree
of the $a$-th row and $\bar{b}$ the Grassmann degree of the $b$-th column.
 We also need
the graded multiplication of matrices,
$$
( K_1 * K_2 )_{ac} \equiv \sum_{b} (-)^{b} {K_1}_{ab}\,{K_2}_{bc}\,.
$$
It is easy to check that the graded matrix transposition relates the 
matrix multiplications of the two types,

\be \lb{grtr}
(K_1 \, K_2 \cdots \, K_n)^ t = K^t_n * K^t_{n-1} * \cdots * K^t_1\,.
\ee
We also need the inversion formula for $\mathrm{L}$-operator (\ref{Lsh}) 
with respect to the 
graded matrix multiplication. Analogously to (\ref{inv}) one can check that
\be \lb{inv'}
\left[\mathrm{L}^t(u) * \mathrm{L}^t(1-u)\right]_{ab} = u(1-u) (-)^{\bar{a}}\delta_{ab}\,.
\ee
The latter relation is valid only at the Casimir operator value equal zero (\ref{cas}).

Now we are ready to prove (\ref{cyc}). We do this in four steps.
First  we multiply the eigenvalue relation by the inverse of the  $\mathrm{L}$-operator
in the first space using  (\ref{inv})
$$
\mathrm{L}_2(u) \cdots \mathrm{L}_n(u) \, M =
\frac{C}{u (1-u)}\, \mathrm{L}_1 (1-u) \, M\,.
$$
Then we perform the graded matrix transposition (\ref{grtr}),
$$
\mathrm{L}^t_n(u)* \cdots *\mathrm{L}^t_2(u) \, M =
\frac{C}{u (1-u)}\, \mathrm{L}^t_1 (1-u) \, M\,.
$$
We multiply from the left by $\mathrm{L}^t_1(u)$ in order to remove the 
matrix operator from r.h.s. using (\ref{inv'}),
$$
\left[ \mathrm{L}^t_1 (u) * \mathrm{L}^t_n(u)* \cdots *\mathrm{L}^t_2(u)\right]_{ab} \, M =
C\, (-)^{\bar{a}}\delta_{ab}\, M\,,
$$
and apply the graded  matrix transposition (\ref{grtr}) once more,
$$
\mathrm{L}_2(u) \cdots \mathrm{L}_n(u) \mathrm{L}_1 (u) \, M = C\, M\, .
$$
In this way  we have succeeded to perform a cyclic shift of the spin chain sites $i \to i+1$, $i+n \equiv i$.

These results are compatible with the 
 well known fact that in $\mathcal{N} = 4$ SYM colour-stripped scattering amplitudes are  
invariant with respect to reflections and cyclic shifts of their legs.

In the analysis of the previous and the following chapters the action of a 
$\mathrm{R}$-operator on both sides of a monodromy
eigenvalue relation is applied repeatedly.
If the sites $i,j$ where $\mathrm{R}_{ij}(0)$ acts non-trivially appear in the
monodromy matrix consecutively in the same order, i.e.
$\cdots\mathrm{L}_i(u) \,\mathrm{L}_j(u)\cdots$, the Yang-Baxter relation
(\ref{RLL}) allows the commutation with this monodromy. 
If the ordering is the opposite one, i.e.$\cdots\mathrm{L}_j(u) \,\mathrm{L}_i(u)\cdots$,
the operator $\mathrm{R}_{ij}(0)$ can be pulled through the monodromy
nevertheless by the following argument:
Turn first to the monodromy eigenvalue relation with the reflected ordering,
which is equivalent by the above results. Act then by $\mathrm{R}_{ij}$ on
this relation, where the commutation is possible by  (\ref{RLL}). 
Return to the relation with the original monodromy by applying the
reflection once more.
By application of cyclicity in an analogous way one can perform the action
of $\mathrm{R}_{n 1}$ on a monodromy eigenvalue relation, where
$\mathrm{L}_1$ is the first and $\mathrm{L}_n$ the last factor in the
monodromy.

The eigenvalue relation and these operations with $\mathrm{R}$
can be extended to the case of inhomogeneous monodromy matrices. Then the 
$\mathrm{R}$-operators  at nonvanishing arguments enter. This case will be
addressed in Sect.~\ref{sec-NH}.

\section{ Inverse Soft Limit }
\setcounter{equation}{0}
\label{sec-ISL}

Rewriting the BCFW relation in the form (\ref{MD}) we
have pulled out one $\mathrm{R}$-operator acting on the amplitudes
 $M_L$ and $M_R$ which are sewed together  by one on-shell leg.
In terms of \cite{AH12} this corresponds to the insertion of the BCFW bridge.
Proceeding further we can represent the amplitudes $M_L$ and $M_R$ in a similar way.
In this way we obtain a sequence of $\mathrm{R}$-operators acting on an
on-shell diagram constructed out of three-point amplitudes.
But three-point amplitudes  can be represented
in $\mathrm{R}$-operator form too as we have shown above,
(\ref{opaMHV3}), (\ref{opMHV3}).
Finally, any amplitude term can be represented as a sequence of $\mathrm{R}$-operators
applied to a product of delta functions corresponding to the external particle states.

Let us establish the connection of this $\mathrm{R}$-operator reconstruction
of  amplitude terms 
 with a well known Inverse Soft Limit (ISL)
iterative procedure proposed in~\cite{ABCCK09} and elaborated in~\cite{Bull10}. It has been applied in~\cite{NW12,BV11} 
to reconstruct BCFW terms for arbitrary tree level amplitudes starting with $3$-point amplitude 
and inserting at each step one additional external state.

We start with the $n-1$-leg amplitude $M_{k,n-1}$ and insert one further particle without 
a change of the Grassmann degree producing $M_{k,n}$. 
One can easily check following the calculation of Subsect.~\ref{sec-3point} 
that in terms of $\mathrm{R}$-operators this takes the form
\be \lb{ISLa}
\mathrm{R}_{n 1} \mathrm{R}_{n \, n-1} \, 
M_{n-1}(1,\cdots,n-1) \,\delta^2(\lambda_{n}) =
\ee
$$
= \frac{\sang{n-1}{1}}{\sang{n-1}{n}\ang{n 1}} \,
M_{n-1}\left(\lambda_1, \frac{(p_1+p_n)\ket{n-1}}{\sang{1}{n-1}},\cdots,
\lambda_{n-1}, \frac{(p_{n-1}+p_n)\ket{1}}{\sang{n-1}{1}} \right) \ = M_n.
$$
Thus two $\mathrm{R}$-operators correspond to an insertion of one additional particle.
Let us check that this procedure is compatible with the monodromy 
eigenvalue condition (\ref{eigenu}).
Assuming that $M_{n-1}$ is an eigenfunction of the $(n-1)$-site monodromy,  
\be \lb{Tn-1} 
\mathrm{T}_{1\cdots n-1}(u) \, M_{n-1} = C_{n-1} \cdot M_{n-1}\,, 
\ee
we see that after multiplication  by a local basic state in the $n$-th site we
produce an eigenfunction of the $n$-th site monodromy (see (\ref{Lvac})),
$$ 
\mathrm{T}_{1\cdots n}(u)  \, M_{n-1} \, \delta^{2}(\lambda_n) = C_{n-1} \cdot M_{n-1} \, \mathrm{L}_n(u)\,
\delta^{2}(\lambda_n) = (u-1) \,C_{n-1} \cdot M_{n-1}\, \delta^{2}(\lambda_n) \,.
$$
In order to entangle the degrees of freedom of the $n$-th particle with the others 
we act with $\mathrm{R}_{n 1}\, \mathrm{R}_{n \, n-1}$ on both sides
of the latter relation to obtain the symmetry condition for $M_n$,
$$
\mathrm{T}_{1\cdots n}(u)  \, M_{n}  = (u-1) \,C_{n-1} \cdot M_{n}\,.
$$
On the basis of cyclicity (\ref{cyc}), reflection relation (\ref{reflect}) 
and the $\mathrm{RLL}$-relation (\ref{RLL})
the operator $\mathrm{R}_{n 1}\, \mathrm{R}_{n \, n-1}$ 
can be pulled through the monodromy.  
More specifically,  first we reflect the chain site ordering 
$1 \, 2 \cdots n \to n\cdots 2\, 1$. 
Then we pull through $\mathrm{R}_{n n-1}$ by means of the $\mathrm{RLL}$-relation, 
perform the cyclic shift 
$n\, {n-1} \cdots 2\, 1 \to 1 \,n \, {n-1} \cdots 2$, pull through $\mathrm{R}_{1n}$ and, 
finally, get back to 
the initial site ordering $1\, 2 \cdots n-1 \,n$ by combining 
a cyclic shift and the reflection.

Notice that for generating an amplitude term the possible $\mathrm{R}$-operator 
actions are restricted also by the condition that the additional
delta function is absorbed by integration over the shifts. This fixes
uniquely the product of $\mathrm{R}$-operators applicable here.

In a similar way we insert the particle of opposite chirality passing from $M_{k,n-1}$ to $M_{k+1,n}$,
$$
\mathrm{R}_{1 n} \mathrm{R}_{n-1 \, n} \,  
M_{k, n-1}(1,\cdots,n-1)\, \delta^{2}(\tilde{\lambda}_{n})\delta^{4}(\eta_n)
=
\frac{\delta^{4}\left([1 \;n-1]\eta_n + [n-1 \; n] \eta_1 + [n 1]\eta_{n-1}\right)}
{[n-1 \;n] [n 1] [n-1 \;1]^3} \cdot
$$
\be \lb{ISLb}
\cdot 
M_{k, n-1}\left(\frac{(p_1+p_n)|n-1]}{[1 \; n-1]},\tilde{\lambda}_1,\cdots,
\frac{(p_{n-1}+p_n)|1]}{[n-1 \; 1]}, \tilde{\lambda}_{n-1} \right)
=  M_{k+1,n}.
\ee
The trivial insertion of the $n$-th particle without interaction 
is compatible with the eigenvalue relation for the $n$-site monodromy (see (\ref{Lvac})),  
$$ 
\mathrm{T}_{1\cdots n}(u)  \, M_{k, n-1} \, \delta^{2|4}(\tilde{\lambda}_n) = C_{n-1} \cdot M_{n-1} \, \mathrm{L}_n(u)\,
\delta^{2|4}(\tilde{\lambda}_n) = u \,C_{n-1} \cdot M_{k,n-1}\, \delta^{2|4}(\tilde{\lambda}_n)\,,
$$
This relation is preserved after pulling 
$\mathrm{R}_{1 n} \mathrm{R}_{n-1 \, n}$ through the monodromy,
$$
\mathrm{T}_{1\cdots n}(u)  \, M_{k+1, n}  = u \,C_{n-1} \cdot M_{k+1, n}\,.
$$

Since arbitrary tree amplitudes and leading singularities of loop corrections can be constructed iteratively 
by means of ISL  we conclude that they can be represented as well as a sequence of $\mathrm{R}$-operators acting on basic state 
$\Omega_{k,n}$ formed by the direct product of $k$ delta functions 
$\delta^{2}(\tilde{\lambda}_i)\delta^{4}(\eta_i)$
and $n-k$ delta functions $\delta^{2}(\lambda_j)$,

In Sect.~\ref{sec-NH} we present analogues of the formulae (\ref{ISLa}), (\ref{ISLb}) 
that allow 
to construct eigenfunctions of inhomogeneous monodromy matrices.

\section{Representations of the Yangian symmetry condition }
\setcounter{equation}{0}
\label{sec-canon}
In this Section we expose some  constructions from~\cite{ChK13} in order to relate them 
with the results obtained above.
Having fixed the representation in algebraic sense, the underlying canonical
variables can be chosen in different ways related by canonical
transformations. We shall distinguish
representations also in this sense which are related to each other  
like the position and momentum representations in Quantum Mechanics.

In the general case of $g\ell(N|M)$ the {\it Yangian symmetric correlators} are defined as 
functions of $n$ points in $N+M$ dimensional super space, where to each point
$\mathbf{x}_k$ one attributes a sign $\kappa_k=\pm$, a dilatation weight $2\ell_k$
and a spectral parameter $u_k$, obeying the monodromy eigenvalue relation
\be \lb{ev}
\mathrm{T}^{\kappa_1,\cdots,\kappa_n }(u_1, \cdots, u_n) \,
M(\mathbf{x}_1,\cdots,\mathbf{x}_n) = C \, M(\mathbf{x}_1,\cdots,\mathbf{x}_n)\,.
\ee
The signature divides the set of $n$ points into the subset $I$ carrying
sign $+$ and $J$ carrying sign $-$.

In the representation used in  \cite{ChK13}
the monodromy matrix has been composed from $\mathrm{L}$-operators of two
types $\mathrm{L}^{\pm}$ depending besides of the spectral parameter on signature,
$$ \mathrm{L}^+(u) = u + \mathbf{p}\otimes \mathbf{x}\,,\,\,\, 
  \mathrm{L}^- (u) = u - \mathbf{x} \otimes \mathbf{p}\,, 
$$
such that in analogy with (\ref{T})
\be \lb{Tpm} 
\mathrm{T}^{\kappa_1,\cdots,\kappa_n }(u_1, \cdots , u_n) = 
\mathrm{L}^{\kappa_1}_1(u_1) \cdots \mathrm{L}^{\kappa_n}_n(u_n) =
\begin{array}{c}
\includegraphics{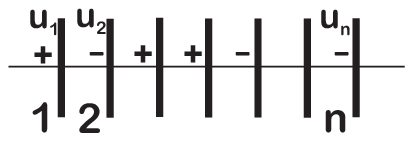}
\end{array} 
\ee
Let us indicate the relation with the notation used in (\ref{L}), 
$\mathrm{L}^{-}(u) = - \mathrm{L}(-u)$.

In this representation the operators
$\mathrm{L}^{\pm}$ and $\mathrm{T}$ act on functions of  $N|M$-component points 
$\mathbf{x}_k,\, k= 1, \cdots, n$ and $\mathbf{p}_k$ act as derivatives.
The simplest solution of the eigenvalue condition (\ref{ev}) playing the role of a
basic state is represented by the constant function 
$M = \Omega \equiv 1 $. 

The
corresponding eigenvalue is 
$$C_0 = \prod_{i \in I} (u_i +1) \prod_{j \in J} u_j .
$$
A general ansatz  is given by the link integral form or equivalently as a sum of 
monomials of fixed dilatation weight 
with respect to each site of the spin chain
$$ 
M = \int \mathrm{d}c \, \phi(c) \exp\biggl(- \sum_{i \in I , j\in J} c_{ij} (\mathbf{x}_i \cdot \mathbf{x}_j )\biggr) =
 \sum b(\lambda) \prod_{i \in I , j\in J} (\mathbf{x}_i \cdot \mathbf{x}_j)^{\lambda_{ij}}\,. 
$$ 
The special feature of this representation is that 
Yangian symmetric correlators are regular functions, i.e. 
working with it we can avoid distributions.

Starting from this representation further ones can be obtained 
by canonical transformations
and in particular 
the representation we have formulated in Sect.~\ref{sec-LR}
as the initial one for this paper.

 Let us first describe the transformation to the {\it uniform} representation.
We apply elementary canonical transformations at the canonical pairs associated
with the points $i \in I$ exchanging there momenta and positions, $\mathbf{x}_i \to -\mathbf{p}_i$ 
and $\mathbf{p}_i \to \mathbf{x}_i$. This
corresponds to Fourier transform of the arguments $\mathbf{x}_i$.
By this transformations 
$$
\mathrm{L}^+_i(u) \longrightarrow \mathrm{L}^{-}_{i}(u)
$$  
and the monodromy (\ref{Tpm}) acquires the form
independent of the signature like (\ref{T}). The information about the signature
carries over to the basic state which acquires the form of the distribution 
\be \lb{OmUni}
\Omega_I = \prod_{i\in I} \delta^{N|M} (\mathbf{x}_i) 
\ee
and other solutions appearing as some operators acting on $\Omega$
keep this information about $I,J$. The general ansatz for this signature 
is now
$$ M  =\int \mathrm{d}c \,\phi(c) \exp\biggl( \sum c_{ij} (\mathbf{p}_i \cdot \mathbf{x}_j) \biggr) \,\Omega_I =
\int \mathrm{d}c \, \phi(c) \prod_{i \in I} \delta^{N|M} \biggl( \mathbf{x}_i + i \sum_{j\in J} c_{ij} \mathbf{x}_j \biggr).
$$
We shall use the  uniform representation in Sect.~\ref{sec-MT} discussing 
Yangian symmetry and Yang-Baxter operators in the specification to  
momentum-twistor variables. 

Now we describe the canonical transformation to spinor-helicity variables.
We separate the $N|M$ components of each point $k= 1,\cdots,n$ in two subsets 
labeled correspondingly by $(\dot{\alpha}, A)$ and $\alpha$
$$ 
\mathbf{x}_{k} = (\, \tilde{\la}_{\alphadot, k} \,,\,\eta_{A,k}\,,\, \la_{\alpha, k} \,) \,.
$$
We have changed here the notations of the coordinates.
The variables matching the spinors appear in the next step.
The canonical pairs of the points $i \in I$ are substituted according to the
elementary canonical transformation 
$$ 
( \, \la_{\alpha, i} \,;\,  \dd_{\alpha, i} \,)
\longrightarrow (\, - \dd_{\alpha, i} \,;\,  \la_{\alpha, i} \,)\,
$$
and the canonical pairs of the points $j\in  J$ are transformed as
$$ 
(\, \tilde{\la}_{\alphadot, j} \,,\,\eta_{A,j}\,;\, \tilde{\dd}_{\alphadot, j} \,,\,\dd_{A,j}\, )
\longrightarrow (\, \tilde{\dd}_{\alphadot, j} \,,\,\dd_{A,j}\,;\, - \tilde{\la}_{\alphadot, j} \,,\,-\eta_{A,j}\, )\,.
$$
The substitutions at $i\in I$ and $j \in J$ lead from $- \mathrm{L}^+(-u)$ and
$-\mathrm{L}^-(-u)$ to one and the same form of the $\mathrm{L}$-operator (\ref{Lsh}) 
such  that the monodromy (\ref{Tpm}) acquires again a form independent of the signature.
The transformation leads also from the original form of the basic state to
$$ \Omega_{k,n} = \prod_I \delta^{2}(\lambda_i) \prod_J
\delta^{2}(\tilde{\lambda}_j)\delta^{4}(\eta_j) $$
where the size of the index set $I$ is $n-k$ and of $J$ is $k$. Here we have
specified the superspace dimensions as $N|M = 4|4$ and the  variable
separation as appropriate for our case.

The eigenfunctions of the monodromy $\mathrm{T}(u_1,\cdots,u_n)$ are in particular
eigenfunctions of its non-diagonal elements with eigenvalue zero.
In the spinor-helicity representation and in the homogeneous case $u_1 = u_2= ...=
u_n = u$ in the expansion of the monodromy in $u$ at the $(n-1)$-st power we have at
non-diagonal positions the total momentum and supercharge
$$ 
\sum_1^n \la_{\alpha} \tilde{\la}_{\dot{\alpha}} \, M = 0 \,\,\,,\,\,\, \sum_1^n \la_{\alpha} \eta_A \, M = 0 \,.
$$
Therefore the eigenfunction $M$ is proportional to the corresponding bosonic and fermionic delta functions,
\be \lb{deltaFact}
M \sim \delta^{4|0} \biggl(\sum_1^n \la_{\alpha} \tilde{\la}_{\dot{\alpha}}\biggr) \, 
\delta^{0|8}\biggl( \sum_1^n \la_{\alpha} \eta_A \biggr)  \,.
\ee
From this point of view the feature of SYM amplitudes that the related MHV amplitude including the above delta
functions can be factorized appears natural.

We have seen that eigenfunctions can be generated by the action on the basic
state by a  sequence of $\mathrm{R}$-operators, if it can be pulled through
the
monodromy according to the procedures described at the end of
Sect.~\ref{sec-cycl}. The integrations involved in the $\mathrm{R}$-operator
actions absorb some of the delta functions in the basic state. In any case
the factor of the momentum and supercharge conservation (\ref{deltaFact})
is left. Amplitude contributions are the ones with no more delta functions
left besides the ones of this factor.

\section{Inhomogeneous monodromy}
\setcounter{equation}{0}
\label{sec-NH}
It can be shown easily that the symmetry conditions and the action
by $\mathrm{R}$-operators  can be considered as limits of the ones 
where the monodromy is inhomogeneous and $\mathrm{R}$-operators enter 
with non-zero arguments. As a sidestep from the
main line of discussion we work out examples of solutions of the deformed
symmetry condition involving 
general inhomogeneous monodromy matrices in spinor-helicity variables.
We introduce the inhomogeneous monodromy matrix constructed from 
$\mathrm{L}$-operators
\be \lb{TNH}
\mathrm{T}_{12\cdots n}(u_1,u_2,\cdots,u_n) = \mathrm{L}_1(u_1)\,\mathrm{L}_2(u_2)\cdots \mathrm{L}_n(u_n) =
\begin{array}{c}
\includegraphics{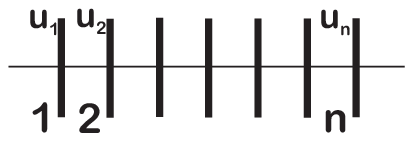}
\end{array}
\ee 
where $u_1,\cdots,u_n$ are spectral parameters and the lower indices refer to
the spin chain sites. 
We will suppress the dependence of the monodromy on the latter when it does not cause misunderstandings.

The Casimir operator $\mathrm{c}$ (\ref{cas}) commutes with the $\mathrm{L}$-operator. 
Consequently $\mathrm{c}_i = \left( \mathbf{p}_i \cdot \mathbf{x}_i \right)$, $i=1,\cdots,N$, 
commute with the monodromy whose eigenfunctions are eigenfunctions of this set of Casimir operators as well.
In the case of homogeneous monodromy (\ref{T}) appropriate for dealing with scattering amplitudes
one has  $\mathrm{c}_i = 0$.
This does not hold in the inhomogeneous case. Therefore the formulae for the inversions of the $\mathrm{L}$-operator 
with respect to standard (\ref{inv}) and graded (\ref{inv'}) matrix products modify
\be \lb{invertC}
\mathrm{L}(u)\,\mathrm{L}\left(1-u-\mathrm{c}\right) = u\left(1-u-\mathrm{c}\right),
\ee
$$
\left[\mathrm{L}^t\left(1-u-\mathrm{c}\right) * \mathrm{L}^t(1-u)\right]_{ab} = u(1-u-\mathrm{c}) (-)^{\bar{a}}\delta_{ab}\,.
$$
As a result the appropriate modification of the cyclicity relation for inhomogeneous monodromy takes the form 
\be \lb{cyclNH}
\mathrm{T}_{12\cdots n}(u_1,u_2,\cdots,u_n) \, M = C \, M \Rightarrow
\mathrm{T}_{2\cdots n 1}(u_2,\cdots,u_n,u_1) \, M = 
C \,\frac{(u_1-1)(u_1+\mathrm{c}_1)}{u_1(u_1+\mathrm{c}_1-1)} \,M\,.
\ee
The reflection relation (\ref{reflect}) 
$1 \, 2\cdots n \to n \cdots 2 \,1$ 
also modifies in an obvious manner by means of (\ref{invertC}).

The formulae of the previous Sections can be recovered from the following
ones taking all spectral parameters equal. Since the basic state $\Omega_{k,n}$ factorizes  it is 
an eigenfunction of the inhomogeneous monodromy as well. 
\subsection{3-point eigenfunctions}
\label{sec-NH3}

In Sect.~\ref{sec-3point} we have reproduced the $3$-point MHV and the anti-MHV
amplitudes  by the action of $\mathrm{R}$-operators (\ref{Rhel})
at $u =0$ on basic states which factorize in the product of local basic states, 
$$
\Omega_{2,3} = \delta^2(\lambda_1) \,\delta^{2}(\tilde{\lambda}_2)\delta^{4}(\eta_2) \,\delta^{2}(\tilde{\lambda}_3)\delta^{4}(\eta_3)\,.
$$
Now we are going to generalize this by taking the Yang-Baxter operator 
$\mathrm{R}(u)$ at arbitrary $u$.
As we shall see shortly  this  leads  to  solutions of the inhomogeneous 
eigenvalue problem.

Let us  start with the parameter deformation of $\mathrm{MHV}_3 = M_{2,3}$  (cf. (\ref{opMHV3})) and show
that
\be \lb{R23R12}
\mathrm{R}_{23}(a) \,\mathrm{R}_{12}(b) \,\Omega_{2,3}
=  \frac{\delta^{4}(p_1+p_2+p_3)\delta^{8}(q_1 + q_2 + q_3)}
{\ang{12}^{1-a} \ang{23}^{1+b} \ang{31}^{1+a-b}} \equiv \mathrm{MHV}_3(a,b)\,.
\ee
In order to prove (\ref{R23R12}) we
take into account (\ref{Rhel}) and obtain
$$
\mathrm{R}_{12}(b) \, \delta^{2}(\lambda_1) \delta^{2|4}(\tilde{\lambda}_2) =
\int \frac{\mathrm{d}z}{z^{1-b}} \delta^2(\lambda_1-z\lambda_2) \delta^{2|4}(\tilde{\lambda}_2+z \tilde{\lambda}_1) =
\delta(\ang{1 2})
\delta^{2|4}\left(\tilde{\lambda}_2+ \frac{\ang{1 3}}{\ang{2 3}} \tilde{\lambda}_1 \right)
\left(\frac{\ang{2 3}}{\ang{1 3}}\right)^{1-b},
$$
where we have rewritten one of the delta functions as
$\delta^2(\lambda_1-z \lambda_2) = \ang{23} \delta(\ang{1 2})\delta( \ang{1 3}-z \ang{2 3})$
projecting its argument on two spinors $\lambda_2$ and $\lambda_3$. 
We could equally well take any other pair of auxiliary spinors without a change in the final answer.
Then we apply one more $\mathrm{R}$-operator
$$
\mathrm{R}_{23}(a)\, \mathrm{R}_{12}(b) \,\Omega_{2,3} =
\left(\frac{\ang{23}}{\ang{1 3}}\right)^{1-b}\int\frac{\mathrm{d}z}{z^{1-a}} \delta(\ang{1 2}-z\ang{1 3})
\delta^{2|4}\left(\tilde{\lambda}_2 + \frac{\ang{1 3}}{\ang{2 3}} \tilde{\lambda}_1 \right)
\delta^{2|4}(\tilde{\lambda}_3 + z \tilde{\lambda}_2) =
$$
$$
= \frac{\ang{2 3}^{1-b}}{\ang{1 2}^{1-a} \ang{1 3}^{1+a-b}}\,
\delta^{2|4}\left(\tilde{\lambda}_2 + \frac{\ang{1 3}}{\ang{2 3}} \tilde{\lambda}_1 \right)
\delta^{2|4}\left(\tilde{\lambda}_3 +  \frac{\ang{1 2}}{\ang{1 3}} \tilde{\lambda}_2 \right) =
\frac{\delta^4(p)\delta^8(q)}{\ang{1 2}^{1-a} \ang{2 3}^{1+b} \ang{3 1}^{1+a-b}}\,.
$$

In a similar way one can prove that the same  deformation of $\mathrm{MHV}_3$ can be obtained applying another
sequence of $\mathrm{R}$-operators
\be \lb{R13R12}
\mathrm{R}_{13}(a) \,\mathrm{R}_{12}(b) \,\Omega_{2,3}
=  \frac{\delta^{4}(p_1+p_2+p_3)\delta^{8}(q_1 + q_2 + q_3)}
{\ang{12}^{1-a} \ang{23}^{1+a+b} \ang{31}^{1-b}}\,.
\ee
 Consequently we have the relation between  (\ref{R23R12}) and (\ref{R13R12})
$$
\mathrm{R}_{23}(a) \,\mathrm{R}_{12}(a+b)\,\Omega_{2,3} = 
\mathrm{R}_{13}(a) \,\mathrm{R}_{12}(b)\,\Omega_{2,3}\,.
$$

Using the representation (\ref{R23R12}) of the $3$-point function $\mathrm{MHV}_3(a,b)$
we are going to check that it is an eigenfunction of the monodromy. 
The $\mathrm{RLL}$-relation (\ref{RLL})  relates unambiguously the parameters $a$ and $b$ in (\ref{R23R12}) 
with the spectral parameters of the monodromy. 
Indeed, one can commute the $\mathrm{R}$-operators through the monodromy
permuting its spectral parameters  in view of the $\mathrm{RLL}$-relation only
in case of appropriate arguments,
\be \lb{TTT}
\begin{array}{c}
\mathrm{T}(u_1,u_2,u_3)\, \mathrm{R}_{23}(u_3-u_2) \,\mathrm{R}_{12}(u_3-u_1) =
\mathrm{R}_{23}(u_3-u_2) \, \mathrm{T}(u_1,u_3,u_2)\, \mathrm{R}_{12}(u_3-u_1) = \\ [0.2 cm ]=
\mathrm{R}_{23}(u_3-u_2)\, \mathrm{R}_{12}(u_3-u_1) \, \mathrm{T}(u_3,u_1,u_2)\,.
\end{array}
\ee
The previous intertwining relation corresponds to the following sequence of permutations on the set of spectral parameters 
$$
u_1 , u_2 , u_3 \rightarrow u_1 , u_3 , u_2 \rightarrow u_3 , u_1 , u_2 \,.
$$
Applying the operator relation (\ref{TTT}) to the basic state $\Omega_{2,3}$ 
and taking into account that $\Omega_{2,3}$ is an eigenfunction of the monodromy, 
$\mathrm{T}(u_3,u_1,u_2)\,\Omega_{2,3} = u_1 u_2 (u_3-1)\, \Omega_{2,3}\,$, 
we have
$$
\mathrm{T}_{123}(u_1,u_2,u_3)\, \mathrm{MHV}_3(u_{32},u_{31}) = 
u_1 u_2 (u_3-1)\cdot \mathrm{MHV}_3(u_{32},u_{31})\,,
$$
where we adopt the shorthand notation $u_{i j} \equiv u_i - u_j$.
Evidently the eigenfunction $\mathrm{MHV}_3(u_{32},u_{31})$ (\ref{R23R12}) is invariant under the simultaneous cyclic shift of 
space  labels $i \to i+1$ (amplitude legs) 
and spectral parameters $u_i \to u_{i+1}$ in agreement with
the cyclicity property of the inhomogeneous 
monodromy (cf. (\ref{cyclNH})).

\vspace{1 em}

In a similar way we proceed with the anti-$\mathrm{MHV}_3$. We start from
another basic state
$$
\Omega_{1,3} = \delta^2(\lambda_1) \,\delta^{2}(\lambda_2) \,\delta^{2}(\tilde{\lambda}_3)\delta^{4}(\eta_3)
$$
and acting on this  by $\mathrm{R}$-operators we obtain the two-parameter
deformation of  anti-$\mathrm{MHV}_3$,
\be \lb{R12R23}
\mathrm{R}_{12}(a) \,\mathrm{R}_{23}(b) \,\Omega_{1,3}=
\frac{\delta^{4}(p_1+p_2+p_3)\delta^{4}([1 2] \eta_3 + [2 3] \eta_1 + [3 1] \eta_2)}{[ 1 2 ]^{1+b} [ 2 3 ]^{1-a} [3 1]^{1+a-b} } 
= \overline{\mathrm{MHV}}_3(a,b) \,.
\ee
In a similar manner the $\mathrm{RLL}$-relation (\ref{RLL})  connects the  parameters $a$ and $b$ in (\ref{R12R23})
to the  spectral parameters of the monodromy resulting in the  eigenvalue relation  
$$
\mathrm{T}(u_1,u_2,u_3)\, \overline{\mathrm{MHV}}_3(u_{21},u_{31}) = 
u_1 (u_2-1) (u_3-1)\cdot \overline{\mathrm{MHV}}_3(u_{21},u_{31})\,.
$$
The expressions for the parameter-deformed amplitudes 
$\mathrm{MHV}_3, \overline{\mathrm{MHV}}_3 $
have been obtained in \cite{FLMPS12} by another method. 

\vspace{1 em}

Now we shall demonstrate on the very simple example of $\mathrm{MHV_3(a,b)}$ 
that the representation of the eigenfunctions of the monodromy
as an excitation of the basic state can be cast into the familiar form of integrals 
over a Grassmannian~\cite{ABCCK09}.
We substitute the $\mathrm{R}$-operators (\ref{Rhel}) 
in the form of integrals over auxiliary parameters in (\ref{R23R12}) and perform
the BCFW shifts in the delta function arguments
$$
\mathrm{R}_{23}(a) \,\mathrm{R}_{12}(b) \,\Omega_{2,3} =
\int \frac{\mathrm{d} z_1}{z_1^{1-a}}\frac{\mathrm{d} z_2}{z_2^{1-b}}
\, \delta^2(\lambda_1 - z_2 \lambda_2 + z_1 z_2 \lambda_3) \delta^{2|4}(\tilde{\lambda}_2 + z_2 \tilde{\lambda}_1)
\delta^{2|4}(\tilde{\lambda}_3 + z_1 \tilde{\lambda}_2) \,.
$$
In order to get rid of bilinear combinations of the auxiliary parameters we perform the variable change
$z_1 \to - \frac{z_3}{z_2}$ resulting in the standard integral over link variables 
$$
\mathrm{MHV}_3(a,b) =
\int \frac{\mathrm{d} z_2\, \mathrm{d} z_3}{z_2^{1+a-b}z_3^{1-a}}
\, \delta^2(\lambda_1 - z_2 \lambda_2 - z_3 \lambda_3) \delta^{2|4}(\tilde{\lambda}_2 + z_2 \tilde{\lambda}_1)
\delta^{2|4}(\tilde{\lambda}_3 + z_3 \tilde{\lambda}_2) \,.
$$
In the next Section we shall show how this  works for the $4$-point eigenfunction.

\subsection{The deformed Inverse Soft Limit } 
In Sect.~\ref{sec-ISL} we have established relations between the $\mathrm{R}$-operator construction of Yangian invariants and
the ISL iterative procedure. Now we  generalize  (\ref{ISLa}) and (\ref{ISLb}) 
to the case of the inhomogeneous monodromy. Calculations similar to the one in the 
previous Subsection
 allow to entangle the local basic state $\delta^2(\lambda_{n})$ with 
the $(n-1)$-point eigenfunctions $M_{n-1}$,
$$
\mathrm{R}_{n 1}(a) \mathrm{R}_{n \, n-1}(b) \, 
M_{n-1}(1,\cdots,n-1)\, \delta^2(\lambda_{n}) =
$$
\be \lb{ISL1}
= \frac{\sang{n-1}{1}^{1-a-b}}{\sang{n-1}{n}^{1-a}\ang{n 1}^{1-b}} \, 
M_{n-1}\left(\lambda_1, \frac{(p_1+p_n)\ket{n-1}}{\sang{1}{n-1}},\cdots,
\lambda_{n-1}, \frac{(p_{n-1}+p_n)\ket{1}}{\sang{n-1}{1}} \right).
\ee
The analogous formula for the opposite chirality local basic state $\delta^{2}(\tilde{\lambda}_{n})\delta^4(\eta_n)$
is
$$
\mathrm{R}_{1 n}(a) \mathrm{R}_{n-1 \, n}(b) \,  
M_{n-1}(1,\cdots,n-1)\, \delta^{2|4}(\tilde{\lambda}_{n})
=
\frac{\delta^{4}\left([1 \;n-1]\eta_n + [n-1 \; n] \eta_1 + [n 1]\eta_{n-1}\right)}{[n-1 \;n]^{1-a} [n 1]^{1-b} [n-1 \;1]^{3+a+b}} \cdot
$$
\be \lb{ISL2}
\cdot 
M_{n-1}\left(\frac{(p_1+p_n)|n-1]}{[1 \; n-1]},\tilde{\lambda}_1,\cdots,
\frac{(p_{n-1}+p_n)|1]}{[n-1 \; 1]}, \tilde{\lambda}_{n-1} \right).
\ee

If $M_{n-1}$ is an eigenfunction of the $(n-1)$-site monodromy 
then the amplitude $M_n$ with one more leg inserted is 
an eigenfunction of the $n$-site monodromy. This can be checked using cyclicity (\ref{cyclNH})
and reflection relations for inhomogeneous monodromy matrices. In the next Subsection
we show how it works in the case of the $4$-point eigenfunction.

\subsection{$4$-point eigenfunction}
The ISL procedure allows to construct higher point eigenfunctions from lower ones. 
In this manner
we add one leg to the deformed $3$-point amplitude (\ref{R23R12}) 
\be \lb{M24R}
M_{2,4}(a,b,c,d) = \mathrm{R}_{1 4}(a) \mathrm{R}_{1 2}(b) \mathrm{R}_{34}(c) \mathrm{R}_{23}(d) \,\Omega_{2,4} 
\ee
where the basic state is now
\be \lb{Om24}
\Omega_{2,4} = \delta^2(\lambda_1) \, \delta^2(\lambda_2) \, 
\delta^{2}(\tilde{\lambda}_3) \delta^{4}(\eta_3) \, \delta^{2}(\tilde{\lambda}_4) \delta^{4}(\eta_4)\,,
\ee
and (\ref{ISL1}) leads to
\be \lb{M24Rexpl}
M_{2,4}(a,b,c,d) =
\frac{\delta^{4}(p_1+\cdots+p_4)\delta^{8}(q_1 + \cdots + q_4)}
{\ang{12}^{1-a} \ang{23}^{1-c} \ang{34}^{1+d} \ang{4 1}^{1-b} \ang{24}^{a+b+c-d}} \,.
\ee
Then we have to adjust the parameters $a,b,c,d$  
for 
$M_{2,4}(a,b,c,d)$ to become  an eigenfunction of the monodromy matrix $\mathrm{T}_{1234}(u_1,u_2,u_3,u_4)$
of the $4$-site spin chain. 

\vspace{1 em}
We start with the $4$-point term obtained by acting three times by $\mathrm{R}$-operators on
the basic state $\Omega_{2,4}$ 
which is the analogue of the cut $4$-point $\mathrm{MHV}$ amplitude
\be \lb{M4cut}
\bcancel{M}_{2,4}(b,c,d) =  \mathrm{R}_{1 2}(b) \mathrm{R}_{34}(c) \mathrm{R}_{23}(d) \,\Omega_{2,4} \,.
\ee
It is an eigenfunction of the monodromy if the parameters are set to the values
$ b= u_{21}, c= u_{43}, d= u_{41}$, 
\be \lb{TMcut}
\mathrm{T}_{1234}(u_1,u_2,u_3,u_4) \,\bcancel{M}_{2,4}(u_{21},u_{43},u_{41}) = u_1 (u_2-1) u_3 (u_4-1)\cdot \bcancel{M}_{2,4}(u_{21},u_{43},u_{41})\,,
\ee
due to the $\mathrm{RLL}$-relation (\ref{RLL}). 
This sequence of $\mathrm{R}$-operators corresponds to the sequence of
permutations
$$
u_1,u_2,u_3,u_4 \rightarrow u_2,u_1,u_3,u_4 \rightarrow u_2,u_1,u_4,u_3 \rightarrow  u_2,u_4,u_1,u_3
$$
on the set of the monodromy parameters. 

In order to construct the full amplitude (\ref{M24R}) we have to consider
the permutation of the operator $\mathrm{R}_{14}$
with the monodromy. However we cannot do this directly in the eigenvalue relation (\ref{TMcut}) by means of
the $\mathrm{RLL}$-relation.
 We have  to transform at first (\ref{TMcut}) by applying reflection and cyclic shift.
We reflect the sequence of spin chain sites in the monodromy (\ref{TMcut})
inverting $\mathrm{L}$-operators by means of (\ref{invertC}) and take into account
the eigenvalues of the Casimir operators (\ref{cas})
on the function $\bcancel{M}_{2,4}(u_{21},u_{43},u_{41})$ 
$$
\mathrm{c}_1 \to u_{21}\,,\, \mathrm{c}_2 \to u_{42}\,,\,\mathrm{c}_3 \to u_{13}\,,\,\mathrm{c}_4 \to u_{34}\,.
$$
Thus (\ref{TMcut}) is transformed into 
$$
\mathrm{T}_{4321}(1-u_3,1-u_1,1-u_4,1-u_2) \,\bcancel{M}_{2,4}(u_{21},u_{43},u_{41}) 
= (u_1-1) u_2 (u_3-1) u_4 \cdot \bcancel{M}_{2,4}(u_{21},u_{43},u_{41})\,.
$$
Next we apply the cyclicity relation (\ref{cyclNH}) to perform the shift of chain sites $4321 \to 1432$,
$$
\mathrm{T}_{1432}(1-u_2,1-u_3,1-u_1,1-u_4) \,\bcancel{M}_{2,4}(u_{21},u_{43},u_{41}) 
= u_1 (u_2-1) (u_3-1) u_4 \cdot \bcancel{M}_{2,4}(u_{21},u_{43},u_{41})\,.
$$
Now we can pull straightforwardly $\mathrm{R}_{14}(u_{32})$ through the monodromy matrix in the previous eigenvalue relation,
$$
\mathrm{T}_{1432}(1-u_2,1-u_3,1-u_1,1-u_4) \,M_{2,4}(u_{32},u_{21},u_{43},u_{41}) = 
u_1 (u_2-1) (u_3-1) u_4 \cdot M_{2,4}(u_{32},u_{21},u_{43},u_{41})\,,
$$
From this eigenvalue relation we can return to the original one
by applying a cyclic shift $1432 \to 4321$,
once more 
and by reflecting the spin chain site ordering $4321 \to 1234$ 
\be \lb{TMHV4}
\mathrm{T}_{1234}(u_1,u_2,u_3,u_4) \,M_{2,4}(u_{32},u_{21},u_{43},u_{41}) = 
u_1 u_2 (u_3-1)(u_4-1) \cdot M_{2,4}(u_{32},u_{21},u_{43},u_{41})\,,
\ee
where according to (\ref{M24Rexpl})
\be \lb{MHV4u}
M_{2,4}(u_{32},u_{21},u_{43},u_{41}) = 
\frac{\delta^{4}(p_1+\cdots+p_4)\,\delta^{8}(q_1 + \cdots + q_4)}
{\ang{12}^{1+u_{23}} \ang{23}^{1+u_{34}} \ang{34}^{1+u_{41}} \ang{4 1}^{1+u_{12}}} =
\begin{array}{c}
\includegraphics{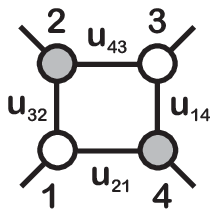}
\end{array}
\ee
The meaning of the latter picture becomes clear by taking into account the results 
of Sect.~\ref{sec-Int} 
where the $\mathrm{R}$-operator is identified with the BCFW bridge.
The expression for the parameter-deformed amplitude 
$\mathrm{MHV}_4 $
has been obtained in \cite{FLMPS12} by other methods.

Thus we have obtained two  nontrivial $4$-point solutions $\bcancel{M}_{2,4}$ (\ref{M4cut}) and $M_{2,4}$ (\ref{MHV4u}) 
of the inhomogeneous monodromy eigenvalue problem. 
Let us note that we could equally well construct $M_{2,4}$ (\ref{MHV4u}) by means of
the BCFW-procedure
following the pattern from Sect.~\ref{sec-BCFW} with appropriate modifications 
due to the inhomogeneity of the monodromy matrix. In that case the parameter
deformed  $4$-point MHV-amplitude
can be obtained by sewing the deformed $\mathrm{MHV}_3$ and anti-$\mathrm{MHV}_3$ each of which satisfies
the eigenvalue relation with the $3$-site inhomogeneous monodromy.

It is evident that we can continue acting by $\mathrm{R}$-operators on the 
basic state $\Omega_{2,4}$
in a way compatible with the monodromy eigenvalue condition. 
Doing so we have to take into account 
cyclicity, reflection and $\mathrm{RLL}$ relations 
allowing to pull the sequence of $\mathrm{R}$-operators 
through the monodromy matrix. It amounts  to fix the dependence of the sequence 
of $\mathrm{R}$-operators 
on the spectral parameters $u_1,\cdots,u_n$ of the inhomogeneous monodromy. Thus we can
excite the basic state $\Omega_{2,4}$
with any number of $\mathrm{R}$-operators producing highly nontrivial 
eigenfunctions of the monodromy.

In the language of on-shell diagrams (see (\ref{RLLpic}))
each $\mathrm{R}$-operator is equivalent to sewing a BCFW bridge.
It introduces one additional integration. 
As we have shown in (\ref{deltaFact}) the delta function 
of the total momentum conservation always factorizes. 
Thus we have to skip $4$ delta functions
in counting the difference of the number of integrations induced by a sequence of $\mathrm{R}$ operators
to  the  number of bosonic delta functions in the basic state. 
Acting three times by a $\mathrm{R}$-operator on $\Omega_{2,4}$ we obtain $\bcancel{M}_{2,4}$ 
which contains one extra bosonic delta function. Acting four times we get $M_{2,4}$ 
with no extra bosonic delta function left. If we proceed  
acting a  fifth time by an
$\mathrm{R}$-operator on the basic state $\Omega_{2,4}$  we come out with one nontrivial integration
left and there is no bosonic delta function to do it trivially.

It is rather evident that applying the ISL-procedure (\ref{ISL1}) and (\ref{ISL2}) and using
the  previous arguments
we can construct the eigenfunctions $M_{k,n}$ of the monodromy 
for any number of legs $n$ and arbitrary Grassmannian degree $4k$.
Following this procedure we have to respect the eigenvalue relation for 
the inhomogeneous monodromy matrix
by specifying 
the arguments of the sequence of $\mathrm{R}$-operators as we have seen above in 
the example of the $4$-point eigenfunction $M_{2,4}$. 
In particular this method enables us to write down the  
$n$-point eigenfunction being a parameter deformation of a $\mathrm{MHV}_n$ 
amplitude and depending on the differences of $n$ spectral parameters $u_1,\cdots,u_n$. 
Moreover we can act 
as many times by $\mathrm{R}$-operators on the basic state $\Omega_{k,n}$ as we want. 
Acting by $\mathrm{R}$-operators $2n-4$ times we result in an analytic function 
multiplied by the ubiquitous
total momentum delta function (\ref{deltaFact}). In the next steps of the procedure nontrivial integrations arise.
In this case as before the monodromy eigenvalue relation determines the dependence 
of the inserted $\mathrm{R}$-operators 
on the spectral parameters $u_1,\cdots,u_n$. 
Let us emphasize that the described method of Yang-Baxter operators 
allows to construct in a rather simple way eigenfunctions that are analogues 
of loop corrections to scattering amplitudes.

\section{Integral  $\mathrm{R}$-operators and Generalized Yang-Baxter relations}
\setcounter{equation}{0}
\label{sec-Int}
In this Section we are going to study integral operators whose kernels 
are eigenfunctions 
of inhomogeneous monodromy matrices, namely the deformed 4-point terms
$\bcancel{M}_{2,4}$ (\ref{M4cut}) and $M_{2,4}$ (\ref{MHV4u}).
In order to show that this is meaningful let us recall the 
relation between the eigenvalue problem for the inhomogeneous monodromy 
and {\it generalized Yang-Baxter relations} introduced in~\cite{ChK13}. 
We  invert $k$ $\mathrm{L}$-operators using (\ref{invertC}) 
together with the fact that the eigenfunctions of the monodromy are also eigenfunctions
of the Casimir operators $\mathrm{c}_{i}$ (\ref{cas}), $i = 1,\cdots,n$,  to 
 rewrite the eigenvalue problem  in the form   
\be \lb{evTrans}
\mathrm{L}_{k+1}(u_{k+1})\cdots \mathrm{L}_{n}(u_{n}) \,
M(\mathbf{x}_1,\cdots,\mathbf{x}_n) =  C\p \,
\mathrm{L}_{k}(u_{k})\cdots \mathrm{L}_{1}(u_{1}) \,M(\mathbf{x}_1,\cdots,\mathbf{x}_n) \,.
\ee
The latter relation can be cast in the form of the intertwining relation  
\be \label{YBgen}
\mathrm{L}_{k+1}(v_{k+1})\cdots \mathrm{L}_{n}(v_{n}) \,
\hat{\mathrm{R}} =  C_R \,
\hat{\mathrm{R}} \, \mathrm{L}_{k}(v_{k})\cdots \mathrm{L}_{1}(v_{1})\, ,
\ee
where the intertwining operator $\hat{\mathrm{R}}$ is considered as
the  integral operator with the kernel 
$M(\mathbf{x}_1,\cdots,\mathbf{x}_n)$
being an eigenfunction of the monodromy
$$
\left[\hat{\mathrm{R}} \cdot F\right] (\mathbf{x}_{k+1},\cdots,\mathbf{x}_{n}) =
\int \mathrm{d} \mathbf{x}_1 \cdots \mathrm{d} \mathbf{x}_k \,
M(\mathbf{x}_1, \cdots , \mathbf{x}_{k} , \mathbf{x}_{k+1} , \cdots , \mathbf{x}_n )\,
F(\mathbf{x}_1, \cdots, \mathbf{x}_k)\,.
$$
The equivalence of (\ref{evTrans}) and (\ref{YBgen}) is established 
by partial integration using (\ref{transp}). 
Thus we see that the eigenvalue problem for the inhomogeneous monodromy is 
definitely related with a Yang-Baxter equation.
Solving the eigenvalue problem we automatically obtain integral Yang-Baxter operators. 
As a particular case the eigenfunction of the $4$-point monodromy 
is the kernel of integral $\mathrm{R}$-operator
which satisfies the ordinary $\mathrm{RLL}$-relation (\ref{RLL}).
The solution of Yang-Baxter $\mathrm{RLL}$-relations by the Yangian
conditions on the corresponding $4$-point kernel has been constructed 
for the case of the $s\ell(2|1)$ symmetry algebra in~\cite{DKK01}.

\vspace{1 em} 
Still working in the spinor-helicity representation  
we are going to take the eigenfunction $\bcancel{M}_{2,4}$ (\ref{M4cut}) 
as the kernel of an integral operator and to show that this operator coincides 
with the $\mathrm{R}$-operator (\ref{Rhel}) which we have used extensively  so far. 
By this calculation  the relation of the  on-shell diagrams~\cite{AH12}   
to the QISM approach will become evident.

Let us define the integral operator $\hat{\bcancel{M}}_{2,4}$ as follows
\be \lb{Mop} 
\left[\hat{\bcancel{M}}_{2,4} F \right] (p_2,\eta_2|p_3,\eta_3) = 
\int \mathrm{d}^4 \eta_1 \mathrm{d}^4 \eta_4 \mathrm{d}^4 p_1 \mathrm{d}^4 p_4 \, \delta(p_1^2) \, \delta(p_4^2) 
\, \bcancel{M}_{2,4}(a,b,c) \, F(p_1,\eta_1|p_4,\eta_4)
\ee
where we integrate over on-shell momenta. 
 The explicit expression for $\bcancel{M}_{2,4}$ (\ref{M4cut}) is 
\be \lb{Mker}
\bcancel{M}_{2,4}(a,b,c) =
\frac{\delta(\ang{12})\delta^{4}(p_1-p_2-p_3+p_4)\delta^{8}(q_1 -q_2-q_3 + q_4)}
{\ang{23}^{1-b} \ang{34}^{1+c} \ang{4 1}^{1-a} \ang{24}^{a+b-c}} = 
\begin{array}{c}
\includegraphics{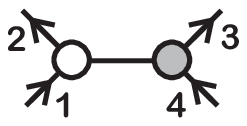}
\end{array}
\ee
We choose the legs $1,4$ to be incoming and $2,3$  outgoing.

\vspace{1 em}

We simplify the integral (\ref{Mop}) in several steps. First we rewrite the delta function in (\ref{Mker}) in the form
$\delta(\ang{12})=[12] \delta\left( (p_1-p_2)^2 \right)$ and consider the
measure of integration
\be \lb{intdelt}
\int \mathrm{d}^4 p_1 \mathrm{d}^4 p_4 \, \delta(p_1^2) \, \delta(p_4^2) \, \delta\left( (p_1-p_2)^2 \right) \, \delta^4(p_1-p_2-p_3+p_4) 
\cdots\,.
\ee
It is clear that in (\ref{intdelt}) only one integration (over $z$) remains  such
that $p_1 = p_2 + z | 2 \rangle [ 3|$ respects the constraints imposed by three delta functions in (\ref{intdelt}).
In order to obtain the integration measure in $z$
 we relax the delta function constraints parameterizing an arbitrary $4$-vector $p_1$ as 
$$
p_1 = p_2 + z | 2 \rangle [ 3| +  z_1 | 2 \rangle [ a| +  z_2 | b \rangle [ 3| +  z_3 | b \rangle [ a|\,.
$$
Here $\tilde{\lambda}_a= [ a|$ and $\lambda_b = |b \rangle$ is a pair of auxiliary spinors. 
At $z_1 = z_2 = z_3 = 0$ the delta function constraints are satisfied.
Performing the change of integration variables from the four-component vector $p_1$ to $z, z_1, z_2, z_3$ we have to calculate
the Jacobian of the transformation $\det K$,
$$
\mathrm{d}^4 p_1 = \det K \cdot \mathrm{d} z \mathrm{d} z_1 \mathrm{d} z_2 \mathrm{d} z_3 \,,\,\,\,
K = \begin{pmatrix} | 2 \rangle [ 3| , | 2 \rangle [ a| , | b \rangle [ 3| , | b \rangle [ a| \end{pmatrix}\,.
$$
For this we apply the reference formula (see~\cite{Dix})
$$
\langle i j \rangle [j l] \langle l m \rangle [m i ] = \frac{1}{2}(s_{i j} s_{l m} - s_{il} s_{j m} + s_{i m} s_{j l})
- 2 i \epsilon_{\mu \nu \rho \sigma} k_i^{\mu} k_j^{\nu} k_l^{\rho} k_m^{\sigma}\,,\,\,\, s_{ij} \equiv 2 k_i \cdot k_j \,,\,\,\,k_i \equiv | i \rangle [i |
$$
and obtain $\det K = \frac{i}{4}\langle 2 b \rangle^2 [a 3]^2 $.
The arguments of the delta functions in (\ref{intdelt}) are
$$
\begin{array}{c}
p_1^2 = \langle b 2\rangle \left( z_2 [2 3] + z_3[2 a] - z z_3 [a 3] + z_1 z_2[a 3]\right)\,,\,\,\,
(p_1-p_2)^2 = [a 3] \langle b 2\rangle \left( z_1 z_2  - z z_3 \right)\,,
\\ [0.2 cm]
p^2_4 = (p_1-p_2-p_3)^2 = [a 3] \left( z_1 \langle 2 3\rangle  + z_3 \langle b 3\rangle + z_1 z_2 \langle b 2\rangle
- z z_3 \langle b 2\rangle \right)\,.
\end{array}
$$
Substituting  to (\ref{intdelt}) and performing sequentially trivial integrations over $z_3$, $z_1$, $z_2$ we
obtain the wanted integration measure
$$
\frac{1}{\langle 2 3\rangle [3 2]} \int \frac{\mathrm{d}z}{z} \cdots\,.
$$
 Notice that the auxiliary spinors $[ a|$ and $|b \rangle$ have  disappeared as it should be.
Further we note the factorizations of the momenta
$|1 \rangle [1| = p_1= |2 \rangle \left( [2| + z [3| \right)$,
$|4 \rangle [4| = p_4 = \left( |3 \rangle - z  |2 \rangle \right) [3|$
 corresponding to the relations between spinors of incoming and outgoing states 
which have the form of a BCFW shift,
\be \lb{spinor41}
|1 \rangle = |2 \rangle \,,\,\,\, |1] = |2] + z |3] \,,\,\,\, |4 \rangle = |3 \rangle - z |2 \rangle \,,\,\,\, |4] = |3]\,.
\ee
The integrations over the Grassmann variables $\eta_1,\,\eta_4$ 
are  done easily since
$$
\delta^{8}(q_1-q_2-q_3+q_4) =
\langle 1 4\rangle^4 
\delta^4 \left( \eta_1 - \eta_2 - z\eta_3 \right)
\delta^4 \left( \eta_4 - \eta_3 \right),
$$
where we take into account (\ref{spinor41}). Simplifying the kernel (\ref{Mker}) by means of 
(\ref{spinor41}) we obtain that the operator (\ref{Mop}) takes the form
$$
\left[\hat{\bcancel{M}}_{2,4} F \right] (p_2,\eta_2|p_3,\eta_3) =
\int \frac{\mathrm{d}z}{z^{1+c}} F(\lambda_2 , \tilde{\lambda}_2 + z \tilde{\lambda}_3 , \eta_2 + z \eta_3 |
\lambda_3 - z \lambda_2, \tilde{\lambda}_3 , \eta_3)
= \mathrm{R}_{32}(-c) F(p_2,\eta_2|p_3,\eta_3).
$$
Thus the operator $\bcancel{M}_{2,4}$ induces the  supersymmetric BCFW shift and 
coincides with the $\mathrm{R}$-operator (\ref{R}). 
This statement clarifies the meaning of the picture form of the $\mathrm{RLL}$-relation (\ref{RLLpic}). 
Now we have shown explicitly that the $\mathrm{R}$-operator does correspond to
the BCFW bridge and has a natural interpretation in terms of on-shell diagrams~\cite{AH12}.
Thus a sequence of $\mathrm{R}$-operators acting on the basic state $\Omega$ corresponds to inserting
successively BCFW bridges producing on-shell diagrams.

\vspace{1 em}

 Let us consider now the integral operator  corresponding to the $4$-point 
eigenfunction (\ref{MHV4u}) related to the $4$-point MHV amplitude,
\be \lb{MFullop}
\begin{array}{c}
\left[\hat{M}_{2,4} F \right] (p_2,\eta_2|p_3,\eta_3) = 
\\ [0.3 cm]
= \int \mathrm{d}^4 \eta_1 \mathrm{d}^4 \eta_4 \mathrm{d}^4 p_1 \mathrm{d}^4 p_4 \, \delta(p_1^2) \, \delta(p_4^2) \, 
M_{2,4}(u_{32},u_{21},u_{43},u_{41}) \, F(p_1,\eta_1|p_4,\eta_4)\,.
\end{array}
\ee
As in the previous case
we choose the legs $1,4$ to be incoming and the legs $2,3$  outgoing.

\vspace{1 em}

In order to rewrite the integral operator (\ref{MFullop}) in a more familiar form we start with the integral over 
the bosonic delta functions
\be \lb{intdelt2}
\int \mathrm{d}^4 p_1 \mathrm{d}^4 p_4 \, \delta(p_1^2) \, \delta(p_4^2) \, \delta^4(p_1-p_2-p_3+p_4) \cdots
\ee
and parametrize the $4$-component momentum $p_1$ by $z_1,z_2,z_3,z_4$,
$$
p_1 = p_2 + z_1 | 2 \rangle [ 3| -  z_1 z_2 | 3 \rangle [ 3| +  z_3 | 3 \rangle [ 2| +  z_4 | 2 \rangle [ 2|\,,
$$
This leads to the transformation of the integration measure
$\mathrm{d}^4 p_1 = - \frac{i}{4} \ang{23}^2[23]^2\, z_1 \cdot \mathrm{d} z_1 \mathrm{d} z_2 \mathrm{d} z_3 \mathrm{d} z_4$.
Taking into account the explicit form of the arguments of the bosonic delta functions
$$
p_1^2 = \langle 2 3\rangle [23] \left( z_1 z_3 + z_1 z_2 z_4 + z_1 z_2\right)\,,\,\,\,
p_4^2 = (p_1-p_2-p_3)^2 = \langle 2 3\rangle [2 3] \left( z_1 z_3  + z_1 z_2 z_4 + z_4 \right)\,,
$$
and performing two integrations in (\ref{intdelt2}) we are left with a double integral
$\int \mathrm{d}z_1\,\mathrm{d}z_2\cdots$
and the relations $z_3 = -z_2(1+z_1 z_2)$, $z_4 = z_1 z_2$ that result in the factorization of
the momenta $p_1$ and $p_4$:
\be \lb{spinor41'}
\begin{array}{lllll}
|1\rangle  = |2\rangle - z_2 |3 \rangle & \,\,\,,\,\,\, & |1] = (1+z_1 z_2)|2] + z_1 |3] &=& |2] + z_1 (|3] + z_2 |2])\,,
\\ [0.2 cm]
|4] = |3] + z_2 |2] & \,\,\,,\,\,\,  & |4\rangle = (1 + z_1 z_2)|3\rangle - z_1 |2 \rangle &= &|3\rangle  - z_1 ( |2 \rangle - z_2 |3\rangle) \,.
\end{array}
\ee
The previous formulae obviously imply two consecutive BCFW shifts. In view of (\ref{spinor41'})
the Grassmann delta function simplifies as follows
$$
\delta^{8}(q_1-q_2-q_3+q_4) =
\langle 1 4\rangle^4 
\delta^4 \left( \eta_1 - \eta_2 - z_1\eta_3 - z_1 z_2 \eta_2 \right)
\delta^4 \left( \eta_4 - \eta_3 - z_2 \eta_2 \right)
$$
supersymmetrizing the BCFW shifts.
Finally we can rewrite the action of the integral operator (\ref{MFullop}) as follows
$$
\int \frac{\mathrm{d}z_1\,\mathrm{d}z_2}{z_1^{1+u_{41}}z_2^{1+u_{23}}} 
F(\lambda_2 - z_2 \lambda_3, \tilde{\lambda}_2 + z_1( \tilde{\lambda}_3 + z_2 \tilde{\lambda}_2 ) , 
\eta_2 + z_1 (\eta_3 + z_2 \eta_2) |
\lambda_3 - z_1 (\lambda_2 - z_2 \lambda_3), \tilde{\lambda}_3 + z_2 \tilde{\lambda}_2, \eta_3 + z_2 \eta_2)\,.
$$
Now it is obvious that two consecutive BCFW shifts in the latter formula 
can be factorized
into the ones of the product of two $\mathrm{R}$-operators (\ref{Rhel})
$$
\left[\hat{M}_{2,4} F \right] (p_2,\eta_2|p_3,\eta_3) 
= \mathrm{R}_{23}(u_{32}) \mathrm{R}_{32}(u_{14}) F(p_2,\eta_2|p_3,\eta_3)\,.
$$
The latter operator relation corresponds to the factorization of the kernel 
\be \lb{factKer}
\begin{array}{c}
\includegraphics{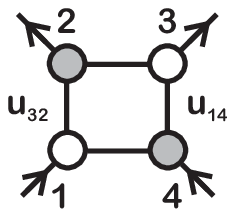}
\end{array} \,\,=
\begin{array}{c}
\includegraphics{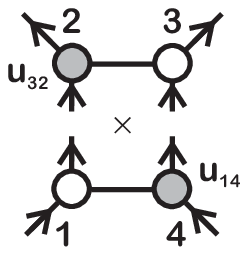}
\end{array}
\ee
In the beginning of this Section
we have recalled  that the eigenvalue problem for
the inhomogeneous $4$-point monodromy matrices
is equivalent to the $\mathrm{RLL}$-relation~\cite{ChK13}. Let us show explicitly how
this works. 
 First we perform the cyclic shift $1234 \to 4123$ in the monodromy eigenvalue problem (\ref{TMHV4})
by means of the cyclicity relation for inhomogeneous monodromy
(\ref{cyclNH}). 
 Then we invert the $\mathrm{L}$-operators 
of the $1$-st and $4$-th sites using (\ref{invertC}),
$$
\mathrm{L}_{2}(u_2)\mathrm{L}_{3}(u_3) \,M_{2,4}(u_{32},u_{21},u_{43},u_{41}) = 
\mathrm{L}_{1}(1-u_3)\mathrm{L}_{4}(1-u_2) \, M_{2,4}(u_{32},u_{21},u_{43},u_{41})\,.
$$ 
Next by partial integration in view of (\ref{transp}) we
rewrite the above equation as an intertwining relation for the integral operator 
$\hat{M}_{2,4}$ (\ref{MFullop}) 
$$
\mathrm{L}_{2}(u_2)\,\mathrm{L}_{3}(u_3) \,\hat{M}_{2,4} = 
\hat{M}_{2,4}\, \mathrm{L}_{2}(u_3)\mathrm{L}_{3}(u_2) \,.
$$
Thus the product $\mathrm{R}_{23}(u)\mathrm{R}_{32}(u)$ respects the $\mathrm{RLL}$-relation
as well as our basic $\mathrm{R}$-operator (\ref{Rhel}). 
This factorization is a direct analogue of the
Yang-Baxter operators factorization used in the construction 
of Baxter $\mathrm{Q}$-operators in~\cite{DKK}.

Let us note that 
if we repeat the previous calculation omitting the cyclic shift $1234 \to 4123$ then 
we arrive at the factorization with respect to the  second unitarity cut in (\ref{factKer}).
The obtained results are  in accordance with Zwiebel's observation~\cite{Zwi} that the
kernels of dilatation operators  coincide with MHV amplitudes. 

Since each $\mathrm{R}$-operator action (\ref{Rhel}) 
contains one integration over an auxiliary parameter $z$
 a sequence of $m$ $\mathrm{R}$-operators is equivalent to a multiple integration over 
$z_1,\cdots,z_m$. Previously in Sect.~\ref{sec-NH3} we have written down 
 such a multiple 
integration in the case of $\mathrm{MHV}_3$
and shown that it corresponds to the link integral representation~\cite{ABCCK09}.
Now we are going to show how this works in the less trivial example of 
the $4$-point eigenfunction $\mathrm{MHV}_4$.
We act by the appropriate sequence of $\mathrm{R}$-operators on the basic state $\Omega_{2,4}$ (\ref{Om24}),
$$
\mathrm{R}_{1 4}(a) \mathrm{R}_{1 2}(b) \mathrm{R}_{34}(c) \mathrm{R}_{23}(d) \,\Omega_{2,4}
=\int 
\frac{\mathrm{d} z_1}{z_1^{1-a}}\frac{\mathrm{d} z_2}{z_2^{1-b}}
\frac{\mathrm{d} z_3}{z_3^{1-c}}\frac{\mathrm{d} z_4}{z_4^{1-d}}\cdot
$$
$$
\cdot \delta^2(\lambda_1 - z_1 \lambda_4 - z_2 \lambda_2) \,
\delta^2(\lambda_2 - z_4 \lambda_3 + z_3 z_4 \lambda_4) \,
\delta^{2|4}(\tilde{\lambda}_3 + z_2 z_4 \tilde{\lambda}_1 + z_4 \tilde{\lambda}_2) \,
\delta^{2|4}(\tilde{\lambda}_3 + z_1 \tilde{\lambda}_1 + z_3 \tilde{\lambda}_3) \,.
$$
The product of delta functions in the previous formula can be rewritten 
 as
$$
\delta^2(\lambda_1 - z_2 z_4 \lambda_3 - (z_1-z_2 z_3 z_4) \lambda_4 ) \,
\delta^2(\lambda_2 - z_4 \lambda_3 + z_3 z_4 \lambda_4) \cdot
$$
$$
\cdot\delta^{2|4}(\tilde{\lambda}_3 + z_2 z_4 \tilde{\lambda}_1 + z_4 \tilde{\lambda}_2)\,
\delta^{2|4}(\tilde{\lambda}_4 + (z_1-z_2 z_3 z_4) \tilde{\lambda}_1 - z_3 z_4 \tilde{\lambda}_2) \,,
$$
where we perform a sequence of variable changes
$$
z_1 \mapsto z_1 + z_2 z_3 z_4 \,\,\,,\,\,\, z_2 \mapsto \frac{z_2}{z_1} \,\,\,,\,\,\, z_3 \mapsto -\frac{z_3}{z_1}
$$
in order to get rid of the quadratic and cubic terms in auxiliary parameters. 
Then we take into account the restriction on the parameters $d = a+b+c$ in (\ref{M24R}) 
induced by the monodromy condition (see (\ref{TMHV4})) and obtain the 
familiar link integral representation
$$
M_{2,4}(a,b,c,a+b+c) = 
\int \frac{\mathrm{d}z_1\,\mathrm{d}z_2\,\mathrm{d}z_3\,\mathrm{d}z_4}
          {(z_1 z_4 - z_2 z_3)^{1-a} z_2^{1-b} z_3^{1-c}}
\prod_{i=1,2}\delta^2(\lambda_i - c_{ji} \lambda_j )\,         
\prod_{j=3,4}\delta^{2|4}(\tilde{\lambda}_j + c_{ji} \tilde{\lambda}_i)\, 
$$
where sums over repeated indices  $i = 1,2$ and $j  = 3,4$ are assumed and the matrix of link variables
is to be identified with the matrix in the integration variables as
$$
||c_{ji}(z)||=
\begin{pmatrix}
c_{31} & c_{32} \\
c_{41} & c_{42}
\end{pmatrix} =
\begin{pmatrix}
z_2 & z_4 \\
z_{1} & z_{3}
\end{pmatrix} \,.
$$
Thus the $\mathrm{R}$-operator construction naturally leads to link integrals over
a Grassmannian
and to on-shell diagrams. 

\section{Monodromy in super momentum-twistor variables}
\setcounter{equation}{0}
\label{sec-MT}

In Sect.~\ref{sec-canon} we have seen that the eigenfunctions of the homogeneous monodromy always
contain the delta functions of total momentum and supercharge conservation (\ref{deltaFact}). 
It is easy to realize that 
the same is true for the inhomogeneous monodromy (\ref{TNH}). Indeed we can perform the shift of 
spectral parameters $u_i \to u_i + u$, $i = 1,\cdots,n$, without  changing
the  eigenfunction $M_n$ since the latter 
depends on the differences $u_{ij}$ of spectral parameters. Then we expand 
the equation in powers of $u$
and follow the argumentation of Sect.~\ref{sec-canon}. 

This motivates to choose variables in such a way
that  this delta function condition is automatically taken into account. 
It is  well-known  that the  (super) momentum twistor variables  introduced 
by Hoges in~\cite{Hod09} have this property.

Super momentum twistors $\mathcal{Z} = (Z,\chi) = (\lambda , \mu , \chi)$ are defined
by the following quasilocal algebraic transformation \cite{MaSk09}
\begin{eqnarray} \lb{MTw}
\tilde{\lambda}_i &=& \frac{\mu_{i-1} \sang{i}{i+1} + \mu_{i} \sang{i+1}{i-1} + \mu_{i+1} \sang{i-1}{i}}{\sang{i-1}{i}\sang{i}{i+1}} \\[0.2 cm]
\eta_i &=& \frac{\chi_{i-1} \sang{i}{i+1} + \chi_{i} \sang{i+1}{i-1} + \chi_{i+1} \sang{i-1}{i}}{\sang{i-1}{i}\sang{i}{i+1}}\,.
\end{eqnarray}
The formulae (\ref{MTw}) are not invertible. 
Arbitrary amplitudes can be represented in the form~\cite{MaSk09,ACC09}
$$
M_{k,n} = M_{2,n} \mathcal{P}_{k-2,n}\,,
$$
where the $\mathrm{MHV}$ amplitude $M_{2,n}$ contains the delta function of the total momentum and supercharge conservation.
The factor $\mathcal{P}_{k-2,n}$ contains all the nontrivial information about the amplitude $\mathrm{N}^{k-2}\mathrm{MHV}_n$.
In fact the function $\mathcal{P}_{k-2,n}$ is a sum of a product of $k-2$ $\mathrm{R}$-invariants.
The $\mathrm{R}$-invariant introduced in~\cite{DHKS08} depends on the variables of five points. 
In~\cite{DrHe08} all tree amplitudes have been constructed solving BCFW relations
in terms of $\mathrm{R}$-invariants in spinor-helicity variables. 
In~\cite{MaSk09} it has been shown that in super momentum twistor space 
the $\mathrm{R}$-invariants are expressed in a particularly simple form.
The simplest $\mathrm{R}$-invariant has the form
\be \lb{lowR-inv}
[1,2,3,4,5] = \frac{\delta^{4}\left(\chi_1 \ang{2345} + \chi_2 \ang{3451}+ \chi_3 \ang{4512}+ \chi_4 \ang{5123}
                         + \chi_5  \ang{1234}\right)}{\ang{1234}\ang{2345}\ang{3451}\ang{4512}\ang{5123}}
\ee
where $\ang{abcd} \equiv \det(Z_a Z_b Z_c Z_d)$ and it is well-known~\cite{DHKS08} 
 that for the $\mathrm{NMHV}$ amplitudes the factor $\mathcal{P}_{1,n}$ is simply 
a sum of such invariants.

Let us choose now the momentum twistors to be the local dynamical variables 
of the spin chain by the
following specifications for coordinates
$\mathbf{x} \to \mathcal{Z}$ and their conjugate momenta 
$\mathbf{p} \to \dd_{\mathcal{Z}}= (\partial_{Z},-\partial_{\chi})$. 
The corresponding $\mathrm{L}$-operator (\ref{L}) 
-- the local building block of the monodromy -- takes the form
\be \lb{Lmt}
\mathrm{L}(u) = u \cdot 1 + \mathcal{Z} \otimes \partial_{\mathcal{Z}} =
\begin{pmatrix}
u \cdot \II + Z \otimes \partial_{Z} & -Z \otimes \partial_{\chi} \\
\chi \otimes \partial_{Z} & u \cdot \II - \chi \otimes \partial_{\chi}
\end{pmatrix}\,.
\ee
Unlike the  spinor-helicity representation (see Sect.~\ref{sec-hel}) 
the $\mathrm{R}$-operator in momentum twistor variables 
intertwining  in the  $\mathrm{RLL}$-relation (\ref{RLL}) products of $\mathrm{L}$-operators 
in the form (\ref{Lmt})  
acts nontrivially only in one of two sites,
\be \lb{Rmt}
\mathrm{R}_{ij}(u) F(\mathcal{Z}_i|\mathcal{Z}_{j}) =
\int \frac{\mathrm{d}z}{z^{1-u}}\,
F(\,Z_{i}- z Z_{j}\,,\, \chi_i - z \chi_j \,|\, Z_j \,,\, \chi_j\,)\,.
\ee
The momentum twistor representation is a particular case of the uniform
representation with coordinates $\mathbf{x}$ identified by $\mathcal{Z}$
and conjugated momenta $\mathbf{p}$ by the corresponding derivatives. 
In this case the basic state has the form (\ref{OmUni})
$$
\Omega_I = \prod_{i \in I} \delta^{4|4}(\mathcal{Z}_i)\,.
$$
It is an eigenfunction of the monodromy built from momentum 
twistor $\mathrm{L}$-operators (\ref{Lmt})
$$
\mathrm{T}(u_1,\cdots,u_n)\,\Omega_I = \prod_{i \in I} (u_i -1) \prod_{j\in J} u_j \cdot \Omega_I\,.
$$
Let us remind that $I \cup J = \{1,2,\cdots,n\}$ and $I \cap J = \varnothing$. 
Indeed from the formulae for
local basic states it follows immediately that
$$
\mathrm{L}(u) \cdot 1 = u \cdot \II \;,\;\;\; 
\mathrm{L}(u) \cdot \delta^{4|4}(\mathcal{Z}) = (u-1) \cdot \delta^{4|4}(\mathcal{Z})\,.
$$
Now acting on $\Omega_{I}$ by a sequence of $\mathrm{R}$-operators (\ref{Rmt}) in a way compatible with
the monodromy condition
we construct more involved eigenfunctions. Let us consider a simple example relevant for scattering amplitudes.
We take the $5$-point monodromy matrix and the basic state 
$\Omega = \delta^{4|4}(\mathcal{Z}_1)$. We 
act on it four times by $\mathrm{R}$-operators in order to absorb  
the four bosonic delta functions.
\be \lb{RinvDef}
\mathrm{R}_{45}(u_{54}) \mathrm{R}_{34}(u_{53}) 
\mathrm{R}_{23}(u_{52}) \mathrm{R}_{12}(u_{51}) \,\delta^{4|4}(\mathcal{Z}_1) =
\ee
$$
= \ang{2345}^{u_{15}} \ang{1345}^{u_{21}} \ang{1245}^{u_{32}} \ang{1235}^{u_{43}} \ang{1234}^{u_{54}}  \, [1,2,3,4,5] \,. 
$$
It is easy to see that this sequence is compatible with the monodromy
condition and matches the permutation 
$$
u_1,u_2,u_3,u_4,u_5 \to u_5,u_1,u_2,u_3,u_4
$$ 
of spectral parameters. Consequently the corresponding eigenvalue is equal to $u_1 u_2 u_3 u_4 (u_5-1)$. 
The calculation in (\ref{RinvDef}) is rather simple and generalizes the one presented in Sec.~\ref{sec-NH3}
from spinors to momentum twistors. For example after the first BCFW shift we rewrite
the bosonic delta function in the form
$$
\delta^{4}(Z_1-z Z_2) = \ang{2345}^3 \delta(\ang{1234}) \delta(\ang{1523}) \delta(\ang{1452}) \delta(\ang{1345} - z \ang{2345})
$$
by projecting  on four different $3$-dimensional planes.

The case relevant for scattering amplitudes corresponds to the homogeneous monodromy 
where all spectral parameters are equal.
In this case the constructed eigenfunction reproduces the $\mathrm{R}$-invariant (\ref{lowR-inv})
\be \lb{Rinv}
[1,2,3,4,5] = \mathrm{R}_{45} \mathrm{R}_{34} \mathrm{R}_{23} \mathrm{R}_{12} \,\delta^{4|4}(\mathcal{Z}_1)\,.
\ee
Once again we see that the involved  highly nonlocal object $[1,2,3,4,5]$ is
obtained by acting in a local way.
As an immediate consequence of the formula (\ref{Rinv}) we conclude that
$\mathcal{P}_{1,n}$ is an eigenfunction of the $n$-site homogeneous monodromy and corresponds to
the eigenvalue $u^{n-1}(u-1)$.

By repeated  $\mathrm{R}$-operator actions one can reconstruct more involved $\mathrm{R}$-invariants 
which are obtained from the simplest one 
(\ref{lowR-inv}) by shifts of its arguments. 
In order to demonstrate how this works let us indicate here the following formula
$$
\begin{array}{c}
\mathrm{R}_{45} \mathrm{R}_{34} \mathrm{R}_{23} \mathrm{R}_{12} \,
\delta^{4|4}(\mathcal{Z}_1)\,F(\mathcal{Z}_2,\mathcal{Z}_3,\mathcal{Z}_4,\mathcal{Z}_5)
 = \\ [0.2 cm]=
[1,2,3,4,5]\,F(\mathcal{Z}_1, \ang{2345}\mathcal{Z}_1 + \ang{3451}\mathcal{Z}_2,
\ang{5123} \mathcal{Z}_4 + \ang{1234}\mathcal{Z}_5,\mathcal{Z}_5)\, ,
\end{array} 
$$
where the function $F$ is assumed to have dilatation weight zero with respect 
to each of its four arguments. 
Consequently the $\mathrm{R}$-operator actions 
reproduce the typical shifts which appear in more involved $\mathrm{R}$-invariants.
 Taking into account 
the previous formula we expect that the explicit solution for all tree amplitudes in terms of super momentum twistors~\cite{BMS,HeMc} 
can be rewritten as sequences of $\mathrm{R}$-operators acting on the basic state $\Omega_I$.

\section{Discussion}
We have formulated Yangian symmetry of super Yang-Mills amplitudes
 in terms of an eigenvalue relation involving the monodromy matrix. 
We have demonstrated that the Quantum Inverse Scattering Method on which this
approach is based provides convenient tools for the calculation and the 
investigation of amplitudes. 
The essential information about the algebraic structure of the symmetry and the particular 
structure of the representation relevant in the application to super Yang-Mills
field theory enters via the choice of the $\mathrm{L}$ matrix. It is the basic
elementary block  from which the monodromy matrix is constructed. Another
important tool is a Yang-Baxter $\mathrm{R}$-operator defined by a standard intertwining
relation with $\mathrm{L}$ matrices. 

We have shown in particular that the proposed Yangian symmetry condition is
compatible with the BCFW iterative calculation. The elementary
three-particle amplitudes obey this symmetry and as a consequence also the
results of the BCFW iteration starting with them.

Solutions of the symmetry condition can be obtained by multiple action with
Yang-Baxter $\mathrm{R}$-operators on basic states. The latter appear in the
spinor-helicity representation as  products of delta distributions
in the spinor variables depending on signature in relation to the
Grassmannian degree. The construction of amplitude contributions by 
$\mathrm{R}$-operators has been demonstrated in a number of examples and the connection
to the Inverse Soft Limit  construction has been explained.

The action of the $\mathrm{R}$-operator induces a particular BCFW shift with an
integral over the shift parameter. 
 The integrations involved in a term with a multiple $\mathrm{R}$ action 
on a basic state  can be transformed into the standard Grassmannian link integral.  
If one prefers instead of this transformation to do the integrals the number of delta
distribution factors present in the basic state is gradually reduced.
In a physical amplitude term all those singular factors are removed
in this way up to the deltas expressing the conservation of total
momentum and supercharge. 

Symmetric amplitude terms can be viewed as integral kernels of
operators acting symmetrically. We have shown that the $\mathrm{R}$-operator
in integral form has the unitarity cut of the four-particle amplitude as its
kernel. By this observation one understands the direct relation between the
$\mathrm{R}$-operator construction and the on-shell diagram approach.

Our approach allows to consider loop contributions not only in
connection with the on-shell diagramm method. We see further ways which
deserve more detailed investigations. The relation of amplitudes to integral
operator kernels allows to generate more symmetric amplitude contributions
from given ones by fusion in terms of integration over the variables of a
number of identified legs as discussed in \cite{ChK13}.
The multiple action by $\mathrm{R}$-operators on a basic state may be
continued after having reproduced the tree amplitude contributions as
considered in examples here. In both ways Yangian invariants are generated
which are naturally related to loop correction of amplitudes. 

It is convenient to consider the Yangian symmetry condition without imposing
any reality constraints related in particular to the signature of space-time.
On the other hand being a tool for generating  amplitude contributions
this  symmetry does not determine completely the physical amplitudes.

The Yangian symmetry condition for amplitude terms involves the
homogeneous monodromy matrix being a product of $\mathrm{L}$ matrices
including a $\mathrm{L}$ factor for each leg with coinciding spectral parameters.
The $\mathrm{R}$-operator appearing in the mentioned construction of amplitude
contributions appears at  zero value of its spectral parameter.
We like to consider the relations for amplitudes as the limiting case of
the ones with the parameters in the $\mathrm{L}$ matrices not all coinciding
(inhomogeneous monodromy matrices) and general values of the spectral parameter argument
of the $\mathrm{R}$-operator. We have shown that this is possible and have provided a
number of examples.

 Our method allows to construct higher point Yangian invariants for
amplitudes with many legs. Presently we do not see straight ways to
compact formulae.

The factor remaining in a general amplitude after the separation of the MHV
amplitude including the momentum and supercharge conservation is known to be
Yangian symmetric as well. Regarding this factor we have formulated the symmetry
condition in terms of momentum twistors and reconstructed by Yang-Baxter 
$\mathrm{R}$-operator
actions the  $\mathrm{R}$-invariant being the basic structure therein.

In this way we have demonstrated how basic and well known features of
SYM amplitudes can be easily derived from Yangian symmetry. 
Relying on the QISM approach   Yangian symmetry has been 
turned from a statement into a practicable working tool.

\section*{Acknowledgement}

The work of D.C. is supported by the Chebyshev Laboratory   
(Department of Mathematics and Mechanics, St. Petersburg State University)
under RF government grant 11.G34.31.0026, by JSC "Gazprom Neft" 
and by Dmitry Zimin's "Dynasty" Foundation. He thanks Leipzig University  
for hospitality and DAAD for support.
The work of S.~D. is supported by RFBR grants
11-01-00570,12-02-91052 and 13-01-12405.

\end{document}